\documentclass[aps,prd,groupedaddress,nofootinbib,amssymb,eqsecnum,epsfig]{revtex4}
\usepackage{graphicx}
\usepackage{bm}
\usepackage{amsmath}
\usepackage{color}
\usepackage{amsfonts}
\usepackage{cases}

\begin{document}
\newcommand{\newc}{\newcommand}

\newc{\be}{\begin{equation}}
\newc{\ee}{\end{equation}}
\newc{\ba}{\begin{eqnarray}}
\newc{\ea}{\end{eqnarray}}
\newc{\aH}{\alpha_{\rm H}}

\title{Absence of solid angle deficit singularities in beyond-generalized Proca theories}

\author{
Lavinia Heisenberg$^{1}$, 
Ryotaro Kase$^{2}$, and 
Shinji Tsujikawa$^{2}$}

\affiliation{
$^1$Institute for Theoretical Studies, ETH Zurich, Clausiusstrasse 47, 8092 Zurich, Switzerland\\
$^2$Department of Physics, Faculty of Science, Tokyo University of Science, 1-3, Kagurazaka,
Shinjuku-ku, Tokyo 162-8601, Japan}

\date{\today}

\begin{abstract}

In Gleyzes-Langlois-Piazza-Vernizzi (GLPV) scalar-tensor theories, which are outside the domain of second-order Horndeski theories, it is known that there exists a solid angle deficit singularity in the case where the parameter $\alpha_{\rm H}$ characterizing the deviation from Horndeski theories approaches a non-vanishing constant at the center 
of a spherically symmetric body. 
Meanwhile, it was recently shown that second-order generalized Proca theories with a massive vector field $A^{\mu}$ can be consistently extended to beyond-generalized Proca theories, which recover shift-symmetric GLPV theories in the scalar limit 
$A^{\mu} \to \nabla^{\mu} \chi$.
In beyond-generalized Proca theories up to quartic-order Lagrangians, 
we show that solid angle deficit singularities are generally absent due to 
the existence of a temporal vector component. 
We also derive the vector-field profiles around a compact object and show that the success of the Vainshtein mechanism operated by vector Galileons is not prevented 
by new interactions in beyond generalized Proca theories.

\end{abstract}


\maketitle

\section{Introduction}

The constantly accumulating observational evidence for an acceleration of the Universe today \cite{SNIa,CMB,BAO} implies that we may require additional dynamical degrees 
of freedom (DOF) to those appearing in standard model of particle physics.
The vacuum energy arising in standard quantum field 
theory can be responsible for the cosmic acceleration, 
but the problem is that the predicted energy scale is 
enormously larger than the observed dark energy 
scale \cite{Weinberg}. 
Instead of resorting to the vacuum energy, there have been 
numerous theoretical attempts for constructing dark 
energy models with some new dynamical DOF \cite{CST}.

A minimally coupled scalar field with a potential is one of 
the simplest dynamical dark energy models \cite{quin}. 
As in Brans-Dicke theory \cite{Brans}, the scalar field can also 
be non-minimally coupled to the Ricci scalar. 
Moreover, we can also think of derivative interactions of 
the scalar field with the Ricci scalar and the Einstein tensor. 
Covariant Galileons \cite{Deffayet,Gali2} have such derivative 
interactions with gravity (see Ref.~\cite{Nicolis} for Minkowski Galileons).
Modified gravitational theories were usually 
constructed to keep the equations of motion up to second order
for avoiding the ghost-like Ostrogradski instability \cite{Ostro}.
Most general scalar-tensor theories with second-order 
equations of motion are known as Horndeski 
theories \cite{Horndeski,Horn2}, 
which accommodate a wide range of single scalar 
dark energy models proposed in the literature. 

Gleyzes-Langlois-Piazza-Vernizzi (GLPV) \cite{GLPV} performed a healthy 
extension of Horndeski theories after expressing the Horndeski Lagrangian 
in terms of ADM scalar quantities with the choice of 
unitary gauge \cite{building}.
Even if the equations of motion can be higher than second order 
in such generalizations, it was shown that the number of dynamical 
scalar DOF remains one in GLPV theories \cite{Hami}.
In the cosmological set-up, the deviation from Horndeski theories 
gives rise to the mixing between sound speeds of the scalar field 
and the matter sector \cite{Gergely,GLPV,Kase14}. 
This property provides tight constraints on some 
dark energy models beyond the Horndeski 
domain \cite{KT14,DKT}.

In GLPV theories, it was shown in Refs.~\cite{conical1,conical2} that 
a solid angle deficit singularity appears in the case where the parameter 
$\alpha_{\rm H}$ characterizing the departure from Horndeski theories approaches a non-zero constant at the center of a spherically symmetric body.
In this case, the Ricci scalar has the dependence $R=-2\alpha_{\rm H}/r^2$ 
around the center of body ($r=0$), so $R$ exhibits the divergence 
at $r=0$. This problem is also related to the breaking of the Vainshtein 
mechanism \cite{Vainshtein} at small radius found 
in Ref.~\cite{Koba} (see also Refs.~\cite{sphe}). 
To avoid these problems, the viable models need to be constructed 
in such a way that $\alpha_{\rm H}$ vanishes for
$r \to 0$ \cite{conical1,Babichev}. 
In such non-singular cases the Vainshtein mechanism can be at work inside 
the solar system, while realizing the successful cosmic expansion 
history \cite{KTD}.

Scalar-tensor theories are not the only possibility for the construction 
of viable dark energy models, but a vector field can be also the source 
for the late-time cosmic acceleration \cite{Barrow,Jimenez1,Jimenez2,TKK,Tasinato,Fleury,Hull,Lagos:2016wyv}. 
For a massive vector-field theory (Proca theory),
the $U(1)$ gauge symmetry is explicitly broken, so that  
the longitudinal mode propagates.
For the construction of theoretically consistent theories, 
the crucial requirement is that self-interactions of the longitudinal mode 
belong to those of Galileon/Horndeski theories.
The condition of second-order equations of motion on curved backgrounds enforces 
the presence of non-minimal derivative couplings with gravity. An interesting subclass of
these type of vector-tensor theories naturally arises in modified gravity theories with Weyl geometries \cite{Jimenez2}.
A systematical classification of derivative vector
self-interactions with three propagating degrees of freedom was carried 
out in Ref.~\cite{Heisenberg}.
The correct number of physical degrees of freedom is
guaranteed by the propagation of second class constraint. Its presence was checked by 
computing the Hessian matrix and the number of its vanishing eigenvalues.
These theories were further investigated in Refs.~\cite{Allys,Jimenez2016}.
In such theories, there are two transverse vector modes and one 
longitudinal scalar besides two tensor polarizations. 
Taking the limit $A^{\mu} \to \nabla^{\mu} \chi$, where $\chi$ is 
a scalar field, the action of generalized Proca theories recovers 
the shift-symmetric action of Horndeski theories \cite{Heisenberg}. 
The dark energy cosmology and spherically symmetric solutions in 
generalized Proca theories were extensively studied 
in Refs.~\cite{Cosmo,Geff,screening,Niz,Minami}.

In Ref.~\cite{HKT} the present authors extended second-order 
generalized Proca theories in such a way that the action of 
new theories can reproduce shift-symmetric GLPV theories 
by taking the scalar limit $A^{\mu} \to \nabla^{\mu} \chi$. 
On the isotropic and anisotropic 
cosmological backgrounds, it was shown that 
beyond-generalized Proca theories do not give rise to 
additional DOF associated with the Ostrogradski ghost 
to those arising in generalized Proca theories \cite{HKT,anisotropy}. 
The absence of extra ghostly DOF up to quartic-order beyond-generalized 
Proca theories was also confirmed in Ref.~\cite{KNY} 
on general curved backgrounds.
Hence it is possible to construct healthy massive 
vector theories even outside the domain of second-order 
generalized Proca theories.

Given the fact that GLPV theories can have the solid angle deficit singularity
at the center of a compact object, we are interested in what happens 
for beyond-generalized Proca theories in the spherically symmetric 
setup. In GLPV theories the source of solid angle deficit singularities is related 
to the geometric modification of the quartic Horndeski Lagrangian. 
In this paper, we consider the quartic-order beyond-generalized Proca 
Lagrangians (whose explicit forms are given in Sec.~\ref{BGPsec})  
and show in Sec.~\ref{BGPconisec} that solid angle deficit singularities 
do not generally appear due to the existence of the temporal 
vector component (see Sec.~\ref{GLPVconisec} for the 
comparison with GLPV theories). 
Thus, beyond-generalized Proca theories have an advantage over 
GLPV theories in that it is not necessary to design
models for avoiding the problem of solid angle deficit singularities.
In Sec.~\ref{Vainsec}, we also study how the new Lagrangian 
affects the vector-field profile around the compact object and show that it does not prevent the success of the Vainshtein mechanism operated by vector-Galileon terms. We conclude in Sec.~\ref{consec}.

\section{Quartic-order beyond-generalized Proca theories 
and the spherical symmetric setup}
\label{BGPsec}

The problem of solid angle deficit singularities in GLPV theories 
can arise in the presence of the quartic Lagrangian $L_4$ beyond 
the domain of Horndeski theories \cite{conical1}. 
The quintic Lagrangian $L_5$ of GLPV theories does not modify
the property of singularities generated by $L_4$ \cite{conical2}. 
Hence the Lagrangian $L_4$ is crucial for the existence of 
solid angle deficit singularities. In this paper, we shall study spherically 
symmetric solutions in beyond-generalized Proca theories  
up to quartic order. 
Besides the vector field $A^{\mu}$ coupled to the Ricci scalar 
$R$, we take into account the matter Lagrangian density ${\cal L}_m$.
The action of quartic-order beyond-generalized Proca theories 
is given by \cite{HKT}
\be
S=\int d^4x \sqrt{-g} \left( {\cal L}_F+ 
\sum_{i=2}^{4} {\cal L}_i
+{\cal L}^{\rm N}_4+{\cal L}_m \right)\,,
\label{action}
\ee
where $g$ is a determinant of the space-time metric 
$g_{\mu \nu}$, and
\ba
{\cal L}_F &=& -\frac14 F_{\mu \nu} F^{\mu \nu}\,,
\label{LF}\\
{\cal L}_2 &=& G_2(X)\,, 
\label{L2}\\
{\cal L}_3 &=& G_3(X) \nabla_{\mu}A^{\mu}\,,
\label{L3}\\
{\cal L}_4 &=& 
G_4(X)R+
G_{4,X}(X) \left[ (\nabla_{\mu} A^{\mu})^2
-\nabla_{\rho}A_{\sigma}
\nabla^{\sigma}A^{\rho} \right]
+\frac12 g_4(X) F_{\mu \nu} F^{\mu \nu}\,,\label{L4}\\
{\cal L}_4^{\rm N}
&=& f_4(X) 
{\cal E}_{\alpha_1 \alpha_2 \alpha_3 \gamma_4} 
{\cal E}^{\beta_1 \beta_2\beta_3\gamma_4}
A^{\alpha_1}A_{\beta_1}
\nabla^{\alpha_2}A_{\beta_2} 
\nabla^{\alpha_3}A_{\beta_3}\,, \label{L4N}
\ea
with $F_{\mu \nu}=\nabla_{\mu}A_{\nu}-\nabla_{\nu}A_{\mu}$, 
and $\nabla_{\mu}$ being the covariant derivative operator.
The functions $G_{2,3,4}, g_{4}, f_{4}$ depend on the quantity  
\be
X \equiv-\frac12A_{\mu} A^{\mu}\,.
\label{defX}
\ee
For the partial derivative with respect to $X$,
we use the notation $G_{i,X} \equiv \partial G_i/\partial X$. 
In ${\cal L}_4$ the non-minimal derivative coupling 
$G_4(X)R$ is required to keep the equations of motion up 
to second order. 
The last term of Eq.~(\ref{L4}) corresponds to the 
intrinsic vector mode, which vanishes by taking the 
scalar limit $A^{\mu} \to \nabla^{\mu}\chi$ \cite{Heisenberg,Allys}. 
The $U(1)$ gauge invariance is explicitly broken by introducing 
the massive Proca Lagrangian $m^2 X$ in ${\cal L}_2$, 
in which case the longitudinal mode of the vector field propagates. 
The number of propagating DOF in second-order generalized 
Proca theories is five on general curved backgrounds (one longitudinal
scalar, two transverse vector modes, and two tensor polarizations). 

The Lagrangian density ${\cal L}_4^{\rm N}$, which contains products 
of the Levi-Civita tensor ${\cal E}_{\alpha_1 \alpha_2 \alpha_3 \gamma_4}$ 
and the covariant derivatives of $A_{\mu}$ up to first order, was constructed 
in a manner analogous to the GLPV extension of scalar-tensor Horndeski theories.
Taking the scalar limit $A^{\mu} \to \nabla^{\mu} \chi$, 
${\cal L}_4^{\rm N}$ recovers the quartic Lagrangian density 
of GLPV theories with the function $f_4$ depending on 
$X=-\nabla_{\mu}\chi \nabla^{\mu}\chi/2$ alone.
Although the new interaction ${\cal L}_4^{\rm N}$ is outside the domain 
of second-order generalized Proca theories, 
it does not give rise to additional ghostly scalar DOF 
on the isotropic cosmological background \cite{HKT}, on
the Bianchi type I background \cite{anisotropy}, and on 
general backgrounds \cite{KNY} . 

To study problems of solid angle deficit singularities and the Vainshtein  
screening inside and outside a compact object, we consider a static and 
spherically symmetric space-time described by the line element 
\be
ds^{2}=-e^{2\Psi(r)}dt^{2}+e^{2\Phi(r)}dr^{2}
+r^{2} d\Omega^2\,,
\label{line}
\ee
where $d\Omega^2=d\theta^{2}+\sin^{2}\theta\,d\varphi^{2}$, 
and the gravitational potentials $\Psi(r)$ and $\Phi(r)$ depend on 
the distance $r$ from the center of symmetry.
We deal with the matter Lagrangian density ${\cal L}_m$ as 
a perfect fluid characterized by the energy-momentum tensor 
$T^{\mu}_{\nu}={\rm diag}(-\rho_m,P_m,P_m,P_m)$. 
We assume that matter is minimally coupled to gravity, such that 
the continuity equation $\nabla^{\mu} T_{\mu \nu}=0$ holds. 
Then, the energy density $\rho_m$ and the pressure $P_m$ obey 
\be
P_m'+\Psi' (\rho_m+P_m)=0\,,
\label{continuity} 
\ee
where a prime represents a derivative with respect to $r$.

The vector field $A^{\mu}$ can be written 
in the form $A^{\mu}=(\phi,A^i)$, where $\phi$ is the temporal 
component and $A^i$ is the three-dimensional vector.
The spatial components are decomposed as $A_i=A_i^{(T)}+\nabla_i \chi$, 
where $A_i^{(T)}$ is the transverse component satisfying 
$\nabla^i A_i^{(T)}=0$ and $\chi$ is the longitudinal scalar. 
On the spherically symmetric background with the coordinate 
$(t,r,\theta,\varphi)$, the $\theta$ and $\varphi$ components of 
$A_i^{(T)}$ vanish, i.e.,  $A_2^{(T)}=0$ and $A_3^{(T)}=0$.
The transverse condition of $A_i^{(T)}$ gives 
the solution $A_1^{(T)}=Ce^{\Phi}/r^2$ \cite{screening}. 
The integration constant $C$ is required to be 0 for the regularity 
at $r=0$, so that $A_1^{(T)}=0$.
Hence the vector field can be expressed as
\be
A^{\mu}=\left( \phi(r),e^{-2\Phi}\chi'(r),0,0 \right)\,.
\label{vecpro}
\ee
On using Eq.~(\ref{vecpro}), the term $X$ 
in Eq.~(\ref{defX}) is decomposed as
\be
X=X_{\phi}+X_{\chi}\,,
\ee
where
\be
X_{\phi}\equiv\frac12e^{2\Psi}\phi^2\,,\qquad
X_{\chi}\equiv-\frac12e^{-2\Phi}\chi'^2\,.
\ee

To derive the background equations of motion, 
we may write the metric (\ref{line}) in a more general form 
$ds^{2}=-e^{2\Psi(r)}dt^{2}+e^{2\Phi(r)}dr^{2}
+r^{2}e^{2\zeta(r)}d\Omega^2$ and express the action 
(\ref{action}) in terms of $\Psi,\Phi,\zeta,\phi,\chi$.
Varying the resulting action with respect to 
$\Psi,\Phi,\zeta,\phi,\chi$ and setting $\zeta=0$ 
in the end, it follows that 
\ba
&&
C_{1}\Psi'^2+\left(C_{2}+\frac{C_{3}}{r}\right)\Psi'+\left(C_{4}+\frac{C_{5}}{r}\right)\Phi'
+C_{6}+\frac{C_{7}}{r}+\frac{C_{8}}{r^2}=-e^{2\Phi}\rho_m
\,,\label{eq00}\\
&&
C_{9}\Psi'^2+\left(C_{10}+\frac{C_{11}}{r}\right)\Psi'
+C_{12}+\frac{C_{13}}{r}+\frac{C_{14}}{r^2}=e^{2\Phi}P_m
\,,\label{eq11}\\
&&
C_{15}\Psi''+C_{16}\Phi''+C_{17}\Psi'^2+C_{18}\Psi'\Phi'+C_{19}\Phi'^2
+\left(C_{20}+\frac{C_{21}}{r}\right)\Psi'+\left(C_{22}+\frac{C_{23}}{r}\right)\Phi'
\notag\\
&&
+C_{24}+\frac{C_{25}}{r}=e^{2\Phi}P_m
\,,\label{eq22}\\
&&
D_{1}(\Psi''+\Psi'^2)+D_{2}\Psi'\Phi'+\left(D_{3}+\frac{D_{4}}{r}\right)\Psi'
+\left(D_{5}+\frac{D_{6}}{r}\right)\Phi'+D_{7}+\frac{D_{8}}{r}+\frac{D_{9}}{r^2}=0
\,,\label{eq0}\\
&&
D_{10}\Psi'^2+\left(D_{11}+\frac{D_{12}}{r}\right)\Psi'+\frac{D_{13}}{r}\Phi'
+D_{14}+\frac{D_{15}}{r}+\frac{D_{16}}{r^2}=0
\,,\label{eq1}
\ea
where the coefficients $C_{1-25}$ and $D_{1-16}$ are given in Appendix A. 
Among the continuity equation (\ref{continuity}) and 
Eqs.~(\ref{eq00})-(\ref{eq1}), five of them are independent.
For instance, it is possible to derive Eq.~(\ref{eq22}) by combining 
other equations of motion.
As a result of going beyond the domain of generalized Proca theories, 
the third-order spatial derivative $\chi'''$ appears in the coefficient $C_{24}$. 

\section{Existence of solid angle deficit singularities in GLPV theories}
\label{GLPVconisec}

We first revisit how solid angle deficit singularities arise in GLPV theories 
to clarify the difference from beyond-generalized Proca theories later.
The quartic GLPV theories can be recovered by taking the 
scalar limit $A^{\mu} \to \nabla^{\mu} \chi$ 
in Eqs.~(\ref{L2})-(\ref{L4N}) (i.e., without the temporal component $\phi$) 
and by allowing the dependence on $\chi$ as well as on 
$X=-\nabla_{\mu} \chi \nabla^{\mu} \chi/2$ 
in the functions $G_{2,3,4}$ and $f_4$.
Then, the Lagrangian densities of GLPV theories 
up to quartic order are given by 
\ba
{\cal L}_2 &=& G_2(\chi,X)\,,\\
{\cal L}_3 &=& G_3(\chi,X) \nabla_{\mu} \nabla^{\mu} \chi\,,\\
{\cal L}_4 &=& 
G_4(\chi, X) R+G_{4,X}(\chi, X) \left[ (\nabla_{\mu}\nabla^{\mu}\chi)^2
-\nabla_{\rho} \nabla_{\sigma} \chi \nabla^{\sigma} \nabla^{\rho} \chi 
\right]\,,\\
{\cal L}_4^{\rm N} &=& 
f_4(\chi, X) 
{\cal E}_{\alpha_1 \alpha_2 \alpha_3 \gamma_4} 
{\cal E}^{\beta_1 \beta_2\beta_3\gamma_4}
\nabla^{\alpha_1} \chi \nabla_{\beta_1} \chi
\nabla^{\alpha_2} \nabla_{\beta_2} \chi 
\nabla^{\alpha_3} \nabla_{\beta_3} \chi\,.
\ea

To quantify the deviation from Horndeski theories, 
we define the parameter\footnote{
The Levi-Civita tensor obeys the normalization 
${\cal E}^{\mu \nu \rho \sigma}{\cal E}_{\mu \nu \rho \sigma}=-4!$, 
which is different from that of Refs.~\cite{conical1,conical2}. 
The definition of $X$ also differs, so we need to 
change $X \to -2X$ and $F_4 \to -F_4$ compared 
to the notations of Refs.~\cite{conical1,conical2}. } 
\be
\alpha_{\rm H}=-\frac{4f_4 X^2}{A_4}\,,
\label{alphaH}
\ee
where 
\be
A_4=2XG_{4,X}-G_4+4f_4X^2\,.
\label{A4}
\ee
The quantity $A_4$ arises after expressing 
the Horndeski Lagrangian in terms of ADM variables 
by choosing the unitary gauge \cite{building,GLPV}. 
In general relativity we have $G_4=M_{\rm pl}^2/2$ and 
$f_4=0$ ($M_{\rm pl}$ is the reduced Planck mass), so 
that $A_4=-G_4=-M_{\rm pl}^2/2$.
One of the simplest models outside the Horndeski 
domain is characterized by constant functions $A_4$ and $-G_4$ with 
$A_4 \neq -G_4$, in which case the terms $4f_4X^2$ as 
well as $\alpha_{\rm H}$ are constants. 
In Ref.~\cite{conical1} it was shown that, for the models with 
non-vanishing $\alpha_{\rm H}$ at $r=0$, there exists the solid angle 
deficit singularity with divergent $R$.

To see how the solid angle deficit singularity arises, we consider 
the model described by 
\be
G_2=X\,,\qquad G_3=0\,,\qquad G_4={\rm constant}\,,
\qquad f_4=\frac{d_4}{X^2}\,,
\label{model1}
\ee
where $d_4$ is a constant. 
In this model the parameter $\alpha_{\rm H}$ is constant, 
which is related to $d_4$, as 
\be
d_4=\frac{\alpha_{\rm H}}{4(1+\alpha_{\rm H})}G_4\,.
\ee
In GLPV theories we have $X_{\phi}=0$ and hence $X=X_{\chi}$.
The non-vanishing coefficients of Eq.~(\ref{eq00}) are given by 
\ba
C_5
&=&-4G_4+16X_{\chi}^2(5f_4+2X_{\chi}f_{4,X})
=-\frac{4G_4}{1+\alpha_{\rm H}}\,,\\
C_8
&=&2(1-e^{2\Phi})G_4-8X_{\chi}^2f_4
=\frac{2G_4}{1+\alpha_{\rm H}}
\left[ 1-e^{2\Phi} (1+\alpha_{\rm H}) \right]\,,
\label{C8GLPV}
\ea
and $C_6=\chi'^2/2$. Let us first derive the solution to $\Phi$
under the assumption that $\chi={\rm constant}$. 
The justification of this assumption will be confirmed later.
Then, Eq.~(\ref{eq00}) reduces to 
\be
-\frac{2G_4}{1+\alpha_{\rm H}} 
\left[ \frac{2\Phi'}{r}-\frac{1-(1+\alpha_{\rm H})
e^{2\Phi}}{r^2} \right]
+e^{2\Phi} \rho_m=0\,.
\label{Phianaeq}
\ee
Assuming that the matter density $\rho_m$ is constant around 
the center of the compact body, 
the solution to Eq.~(\ref{Phianaeq}) is given by 
\be
\Phi=-\frac12 \ln \left[ 1+\alpha_{\rm H}
-\frac{(1+\alpha_{\rm H})\rho_m r^2}{6G_4} \right]\,,
\label{Phianaso}
\ee
where the integration constant $C$, which appears as 
the form $C/(2G_4r)$ in the logarithmic term of 
Eq.~(\ref{Phianaso}), has been set to 0 for the regularity 
of $\Phi$ at $r=0$. 
In the limit that $r \to 0$, there exists the non-vanishing 
constant $-(1/2)\ln (1+\alpha_{\rm H})$ in $\Phi$ for 
$\alpha_{\rm H} \neq 0$. 
As we will see below, this is the source for solid angle 
deficit singularities at $r=0$.

The solution (\ref{Phianaso}) was obtained by assuming that 
$\chi$ is constant. In the following we iteratively derive the 
solutions to $\Phi,\Psi,\chi$ by expanding them around the 
center of the spherically symmetric body, as
\be
\Phi(r) = \Phi_0+\sum_{i=1}^{\infty} \Phi_i r^i\,, \qquad
\Psi(r) =\Psi_0+\sum_{i=1}^{\infty} \Psi_i r^i\,,\qquad
\chi(r) = \chi_0+\sum_{i=2}^{\infty} \chi_i r^i\,,
\label{graexpand}
\ee
where $\Phi_0$, $\Phi_i$, $\Psi_0$, $\Psi_i$, $\chi_0$, $\chi_i$
are constants. The field $\chi(r)$ should satisfy the regular boundary 
condition $\chi'(0)=0$, so the term $\chi_1 r$ is absent.
The matter density $\rho_m(r)$ can be also expanded around $r=0$, 
but the variation of $\rho_m(r)$ does not affect the discussion of 
solid angle deficit singularities \cite{conical1,conical2}.
Hence it is sufficient to consider the case of constant $\rho_m$.
In this case, Eq.~(\ref{continuity}) is integrated to give
\be
P_m(r)=-\rho_m+\rho_s e^{-\Psi(r)}\,,
\label{Pmso}
\ee
where $\rho_s$ is a constant.  

The equations of motion in shift-symmetric GLPV theories can be derived by 
setting $\phi=0$ in Eqs.~(\ref{eq00})-(\ref{eq1}). 
Since Eq.~(\ref{eq0}) is decoupled from the system, 
three of Eqs.~(\ref{eq00})-(\ref{eq1}) are independent.
Substituting Eqs.~(\ref{graexpand})-(\ref{Pmso}) into 
Eqs.~(\ref{eq00}), (\ref{eq11}), and (\ref{eq1}) 
for the model (\ref{model1}),  
we can iteratively obtain the following solutions
\ba
\Phi(r) &=& -\frac12 \ln \left( 1+\aH \right) 
+\frac{\rho_m}{12G_4}r^2+{\cal O}(r^4)\,,
\label{Phiso} \\
\Psi(r) &=& \Psi_0-\frac{2\rho_m e^{\Psi_0}-3\rho_s}
{24G_4 e^{\Psi_0}}r^2
+{\cal O}(r^4)\,,
\label{Psiso}\\
\chi(r)&=&\chi_0\,.
\label{phiso}
\ea
Thus, the assumption that $\chi(r)={\rm constant}$ used for 
the derivation of the solution (\ref{Phianaso}) is justified. 
In fact, expansion of the solution (\ref{Phianaso}) around $r=0$ 
leads to Eq.~(\ref{Phiso}).
On using Eqs.~(\ref{Phiso}) and (\ref{Psiso}), 
the Ricci scalar reads
\be
R=-\frac{2\aH}{r^2}
+\frac{(4\rho_m e^{\Psi_0}-3\rho_s)(1+\aH)}
{2G_4e^{\Psi_0}}+{\cal O}(r^2)\,.
\ee
If the parameter $\aH$ does not vanish at $r=0$, there is the solid angle deficit 
singularity with divergent $R$. 

The model (\ref{model1}) is the simplest one in which the solid angle deficit singularity 
is present. In Refs.~\cite{conical1,conical2}, the authors considered 
more general cases in which additional functions $G_2, G_3,G_4, f_4$ to the model (\ref{model1}) 
are taken into account. Provided $\alpha_{\rm H} \neq 0$ at $r=0$, 
it was shown that these additional contributions do not modify 
the existence of solid angle deficit singularities.
To avoid the appearance of solid angle deficit singularities, we need to construct
models in which $\alpha_{\rm H}$ vanishes at $r=0$. 
For example, the model with the functions $f_4={\rm constant}$ and 
$G_4=M_{\rm pl}^2F(\chi)/2+b_4X^2$ ($F(\chi)$ is a function of $\chi$ 
and $b_4$ is a constant) gives $\alpha_{\rm H}=0$ at $r=0$ due to 
the boundary condition $\chi'(0)=0$, while $\alpha_{\rm H}$ does 
not vanish for $r>0$. 
In such models, not only the problem of solid angle deficit singularities is absent, 
but also the Vainshtein screening can be at work outside the
compact body \cite{conical1,conical2}.

\section{Absence of solid angle deficit singularities in 
beyond-generalized Proca theories}
\label{BGPconisec}

The appearance of solid angle deficit singularities in GLPV theories is intrinsically 
related to the non-vanishing term $-(1/2)\ln (1+\aH)$ in Eq.~(\ref{Phiso}). 
In fact, the regularities of $R$ and the curvature scalars 
$R_{\mu \nu}R^{\mu \nu}$, $R_{\mu \nu \rho \sigma}R^{\mu \nu \rho \sigma}$ 
generally require the three conditions $\Phi_0=0$, $\Phi_1=0$, and 
$\Psi_1=0$ in Eq.~(\ref{graexpand}) \cite{conical2}. 
In other words, if the gravitational potentials are expanded 
around $r=0$ as
\ba
\Phi(r) &=& \Phi_2 r^2+{\cal O} (r^4)\,,\label{Phire} \\
\Psi(r) &=& \Psi_0+\Psi_2 r^2+{\cal O} (r^4)\,, \label{Psire}
\ea
the solid angle deficit singularity is absent. 

In what follows, we show that solid angle deficit singularities do not arise 
in beyond-generalized Proca theories (\ref{action}) by virtue of 
the existence of the temporal vector component $\phi$ 
besides the longitudinal component $\chi$.
To characterize the deviation from 
second-order generalized Proca theories, we define the 
quantity analogous to Eq.~(\ref{alphaH}), i.e., 
\be
\alpha_{\rm P}=\frac{4f_4X^2}
{G_4-2XG_{4,X}-4f_4X^2}\,.
\ee
The regular boundary conditions of the vector field at 
$r=0$ are given by 
\be
\phi'(0)=0\,,\qquad \chi'(0)=0\,.
\label{boundary}
\ee
Since $\phi$ and $\chi$ stay nearly constants around $r=0$, 
we have that 
\be
X_{\chi} (r \to 0)=0\,,\qquad 
X_{\phi} (r \to 0)=\frac{1}{2} e^{2\Psi}\phi^2
={\rm constant}\,.
\label{Xlimit}
\ee
The key difference from GLPV theories is that, provided $\phi \neq 0$, 
$X$ approaches the non-vanishing constant $X_{\phi}$
as $r \to 0$.

To understand the difference from GLPV theories, 
we begin with the following model
\be
G_2=b_2X\,,\qquad
G_3=0\,,\qquad 
G_4={\rm constant}\,,\qquad 
g_4={\rm constant}\,,\qquad 
f_4=\frac{\alpha_{\rm P}}{4(1+\alpha_{\rm P})} 
\frac{G_4}{X^2}\,,
\label{model2}
\ee
where $b_2$ and $\alpha_{\rm P}$ are constants. 
This is analogous to the model (\ref{model1}) in which 
solid angle deficit singularities arise in GLPV theories.
Then, the coefficients $C_5$ and $C_8$ in Eq.~(\ref{eq00}) 
are given, respectively, by 
\ba
C_5
&=&-4G_4+\frac{4\alpha_{\rm P}G_4}{1+\alpha_{\rm P}}
\frac{X_{\chi}(3X_{\phi}+X_{\chi})}
{(X_{\phi}+X_{\chi})^2} \,,
\label{C5b}\\
C_8
&=&2(1-e^{2\Phi})G_4-\frac{2\alpha_{\rm P}G_4}
{1+\alpha_{\rm P}} \frac{X_{\chi}}{X_{\phi}+X_{\chi}}\,.
\label{C8b}
\ea
On account of the relation (\ref{Xlimit}), the last two terms
of Eqs.~(\ref{C5b}) and (\ref{C8b}) vanish for $r \to 0$,
so that $C_5 \to -4G_4$ and $C_8 \to 2(1-e^{2\Phi})G_4$. 
Unlike GLPV theories, the coefficients $C_5$ and $C_8$ do not 
contain the $\alpha_{\rm P}$ term. 
In GLPV theories the term $\alpha_{\rm H}$ 
arising in the square bracket of $C_8$ in Eq.~(\ref{C8GLPV}) 
is the main source for the existence of solid angle deficit singularities. 
In beyond-generalized Proca theories, 
the absence of the terms $\alpha_{\rm P}$ 
in $C_5$ and $C_8$ implies that 
solid angle deficit singularities
may not arise at $r=0$.

To see explicitly the absence of solid angle deficit singularities 
for the model (\ref{model2}), we expand the 
temporal vector component around $r=0$, as
\be
\phi(r) = \phi_0+\sum_{i=2}^{\infty} \phi_i r^i\,,
\label{phiexpansion}
\ee
where $\phi_0,\phi_i$ are constants. 
The gravitational potentials $\Phi, \Psi$ and the longitudinal scalar 
$\chi$ are expanded in the same way as Eq.~(\ref{graexpand}).
For constant density $\rho_m$ around $r=0$, the matter 
pressure $P_m$ is given by Eq.~(\ref{Pmso}). 
Unlike GLPV theories in which $\phi$ exactly vanishes, 
Eq.~(\ref{eq0}) plays an important role to determine 
the field profile of $\phi$.
Substituting Eqs.~(\ref{graexpand}) and (\ref{phiexpansion}) with 
Eq.~(\ref{Pmso}) into Eqs.~(\ref{eq00}), (\ref{eq11}), (\ref{eq0}), (\ref{eq1}) 
and solving them iteratively, we obtain the following solutions
\ba
&&
\Phi(r)=\frac{2\rho_m+b_2e^{2\Psi_0}\phi_0^2}{24G_4}r^2
+{\cal O}(r^4)\,,\label{sol1}
\\
&&
\Psi(r)=\Psi_0
-\frac{2\rho_m-3e^{-\Psi_0}\rho_s
-2b_2e^{2\Psi_0}\phi_0^2}{24G_4}r^2
+{\cal O}(r^4)
\,,\label{sol2} \\
&&
\phi(r)=\phi_0
+\frac{\phi_0}{6}\left( \frac{b_2}{1-2g_4}
+\frac{2\rho_m-3e^{-\Psi_0}\rho_s
-2b_2e^{2\Psi_0}\phi_0^2}{2G_4} \right) r^2
+{\cal O}(r^4)
\,,\\
&&
\chi(r)=\chi_0\,,
\ea
which are expanded up to second order in $r$. 
Compared to Eq.~(\ref{Phiso}), there is no constant term containing 
$\alpha_{\rm P}$ on the r.h.s. of Eq.~(\ref{sol1}). 
The solutions (\ref{sol1}) and (\ref{sol2}) are of the same forms as 
Eqs.~(\ref{Phire}) and (\ref{Psire}) respectively, so we do not have 
solid angle deficit singularities at $r=0$. 
In fact, the Ricci scalar is given by 
\be
R=\frac{4\rho_m-3e^{-\Psi_0}\rho_c
-b_2e^{2\Psi_0}\phi_0^2}{2 G_4}+{\cal O}(r)\,,
\ee
which is finite at $r=0$.

So far we have shown the absence of solid angle deficit singularities 
for the model described by Eq.~(\ref{model2}), but 
this property generally holds even without restricting 
the functional forms of  $G_{2,3,4}, g_4, f_4$. 
To prove this, we use 
the relations (\ref{Xlimit}) with the boundary 
conditions (\ref{boundary}).
Then, around $r=0$, Eqs.~(\ref{eq00}) and (\ref{eq11}) 
reduce, respectively, to 
\ba
&& 
-2(G_4-2X_{\phi}G_{4,X}) \left( \frac{2\Phi'}{r}
-\frac{1-e^{2\Phi}}{r^2} \right)
+C_1 \Psi'^2+C_6+e^{2\Phi} \rho_m \simeq 0\,,
\label{Phiap}\\
&& 4(G_4+2X_{\phi}G_{4,X}) 
\frac{\Psi'}{r}+2(1-e^{2\Phi})\frac{G_4}{r^2} 
+C_9 \Psi'^2+C_{12}
+ e^{2\Phi} \left( \rho_m-\rho_s e^{-\Psi} \right) \simeq 0\,,
\label{Psiap}
\ea
where $C_1,C_6,C_9,C_{12}$ are constants. 
Substituting the expanded gravitational potentials (\ref{graexpand}) 
into Eq.~(\ref{Phiap}), it follows that the two terms 
$2(1-e^{2\Phi_0})(G_4-2X_{\phi}G_{4,X})/r^2$ and 
$-4\Phi_1 (1+e^{2\Phi_0})(G_4-2X_{\phi}G_{4,X})/r$ 
are required to vanish. 
As long as  $G_4 \neq 2X_{\phi}G_{4,X}$, we have 
\be
\Phi_0=0\,, \qquad \Phi_1=0\,.
\label{Phi01}
\ee
{}From Eq.~(\ref{Psiap}) we also find that there is one 
term $4(G_4+2X_{\phi}G_{4,X})\Psi_1/r$ that must vanish.
Provided $G_4 \neq -2X_{\phi}G_{4,X}$, we obtain 
\be
\Psi_1=0\,.
\ee
Then, the gravitational potentials $\Phi(r)$ and $\Psi(r)$ reduce 
to the forms (\ref{Phire}) and (\ref{Psire}) around $r=0$, respectively, 
so the solid angle deficit singularity is absent.
The specific case satisfying $G_4=2X_{\phi}G_{4,X}=-2X_{\phi}G_{4,X}$ 
corresponds to $G_4=0$, so this does not correspond to a realistic situation 
where the general relativistic behavior is recovered.
Hence the absence of solid angle deficit singularities is very generic 
in quartic-order beyond-generalized Proca theories.

\section{Vainshtein mechanism}
\label{Vainsec}

In this section we discuss how the new term 
${\cal L}_4^{\rm N}$ in beyond-generalized 
Proca theories affects the screening mechanism 
in second-order generalized Proca theories.
For concreteness, we study the 
theories described by 
\be
G_2(X)=m^2X\,,\qquad
G_3(X)=\beta_3 X\,,\qquad
G_4(X)=\frac{M_{\rm pl}^2}{2}+\beta_4 X^2\,,\qquad 
g_4(X)=0\,,\qquad
f_4(X)=d_4 X^n\,,
\label{L4lag}
\ee
where $m^2$, $\beta_3$, $\beta_4$, $d_4$, $n$ are constants. 
We assume that $|n|$ is of the order of unity.
When $d_4=0$, the theories reduce to vector 
Galileons \cite{Heisenberg}. 
In principle we can consider more general functions of 
$G_{2,3,4},g_4$, but the above theories are sufficient to 
address the problem of how the new interaction $f_4(X)$ 
affects the Vainshtein mechanism mediated by vector 
Galleons (see Refs.~\cite{scava} for the Vainshtein mechanism 
mediated by scalar Galileons).

For the functions (\ref{L4lag}) the vector-field equations
of motion (\ref{eq0}) and (\ref{eq1}) are given, respectively, by
\ba
& &
\frac{1}{r^2} \frac{d}{dr} \left( r^2\phi' \right)
-e^{2\Phi}m^2\phi
+2\phi \left( \Psi''+\Psi'^2-\Psi'\Phi' \right)
+\left( 3\phi'+\frac{4\phi}{r} \right)\Psi'-\phi'\Phi'
-\beta_3 \phi \left[ \frac{1}{r^2}\frac{d}{dr}
\left( r^2\chi' \right)+\left( \Psi'-\Phi' \right)\chi' \right]  \nonumber \\
& &
-\frac{2\beta_4e^{-2\Phi}\phi}{r^2}
\left[ 4r\chi'\chi'' +e^{2\Psi+2\Phi}\phi^2
\left( e^{2\Phi}-1+2r\Phi' \right)-\chi'^2
\left\{ e^{2\Phi}-3+2r(3\Phi'-2\Psi') \right\} \right] \nonumber
\\
& & 
-\frac{2^{2-n}d_4e^{-2\Phi}\phi \chi'}{r^2} 
\left( e^{2\Psi}\phi^2-e^{-2\Phi}\chi'^2 \right)^n 
\left[2(2+n)r \chi''+\{ 2+n-r\Phi' (5+2n)
+r\Psi' (3+2n) \}\chi' \right]
=0\,,
\label{F0full}\\
& &
m^2\chi'+\beta_3 \left[ e^{2\Psi} (\phi \phi'+\phi^2 \Psi')
+e^{-2\Phi}\chi'^2 \left( \frac{2}{r}+\Psi' \right)
\right] \nonumber \\
& &
+\frac{2\beta_4\chi'}{r} \biggl[
e^{2\Psi}\frac{\phi^2}{r} (1-e^{-2\Phi})+2e^{2\Psi-2\Phi}
(2\phi\phi'+\phi^2 \Psi')-e^{-2\Phi} \frac{\chi'^2}{r}
(1-3e^{-2\Phi}-6r\Psi'e^{-2\Phi}) \biggr] \nonumber \\
& &
+\frac{2^{2-n}d_4e^{-4\Phi}\chi'}{r^2} 
\left( e^{2\Psi}\phi^2-e^{-2\Phi}\chi'^2 \right)^n 
\left[ e^{2\Psi+2\Phi} \phi \{ 2(2+n)\phi'+(2n\Psi'+3\Psi'-\Phi')\phi \}r
+(2+n)(1+2r \Psi')\chi'^2 \right] \nonumber \\
& &
=0\,.
\label{F1full}
\ea
The vector mass squared $m^2$ can be either positive or negative. 
Applying the above model to dark energy, it is natural to consider the 
mass scale $|m|$ of the order of $H_0 \approx 10^{-33}$~eV \cite{Cosmo}. 
For the study of spherically symmetric solutions in the solar system, 
we take the limit $m^2 \to 0$ in the following discussion.

We consider a compact body with the radius $r_*$ and the constant 
density $\rho_0$, i.e., $\rho_m(r)=\rho_0$ for $r<r_*$ 
and $\rho_m(r)=0$ for $r>r_*$.
We also employ the approximation of weak gravity under which the 
Schwarzschild radius $r_g \approx \rho_0 r_*^3/M_{\rm pl}^2$ 
of the body is much smaller than $r_*$, i.e., 
\be
\Phi_* \equiv \frac{\rho_0 r_*^2}{M_{\rm pl}^2} \ll 1\,.
\ee
The general relativistic solutions to $\Phi$ and $\Psi$ 
in the absence of the vector field can 
be derived by setting $G_2=G_3=0$, 
$G_4=M_{\rm pl}^2/2$, and $\phi=\chi'=0$ 
in Eqs.~(\ref{eq00}) and (\ref{eq11}). 
They are given, respectively, by \cite{screening}  
\be
\Phi_{\rm GR}=\frac{\Phi_*}{6} \frac{r^2}{r_*^2}\,,\qquad 
\Psi_{\rm GR}=\frac{\Phi_*}{12} \left( \frac{r^2}{r_*^2}-3 \right)\,,
\label{grain}
\ee
for $r<r_*$, and
\be
\Phi_{\rm GR}=\frac{\Phi_* r_*}{6r}\,,\qquad 
\Psi_{\rm GR}=-\frac{\Phi_* r_*}{6r}\,,
\label{graout}
\ee
for $r>r_*$. In the presence of the vector field coupled to 
gravity, these solutions are subject to modifications. 
As long as the Vainshtein mechanism is at work, 
the corrections to $\Phi_{\rm GR}$ and 
$\Psi_{\rm GR}$ should be small. 
For the theories given by the functions (\ref{L4lag}),
we shall estimate the corrections to Eqs.~(\ref{grain}) 
and (\ref{graout}).

\subsection{$\beta_3=0$}

If $\beta_3=0$, then Eq.~(\ref{F1full}) admits the solution 
\be
\chi'=0\,,
\label{chi0}
\ee
for which the longitudinal vector component $\chi$ stays constant.
In this case the last line on the l.h.s. of Eq.~(\ref{F0full}) vanishes 
identically, so the Lagrangian density ${\cal L}_4^{\rm N}$ 
does not give rise to any contribution to the profile of $\phi$. 
Hence the solution of $\phi$ is similar to the one already 
derived in Ref.~\cite{screening} for $\chi'=0$.

Following Ref.~\cite{screening}, we search for solutions where $\phi$ 
stays nearly constant around a constant $\phi_0$, such that 
\be
\phi(r)=\phi_0+f(r)\,,\qquad 
|f(r)| \ll |\phi_0|\,,
\label{phifr}
\ee
where $f(r)$ is a function of $r$. 
For the integration of Eq.~(\ref{F0full}), the 
temporal component $\phi$ is approximated as $\phi_0$. 
Under the approximation of weak gravity, we neglect the terms 
like $\phi'\Phi'$ relative to the first contribution on the 
l.h.s. of  Eq.~(\ref{F0full}).
We also substitute the gravitational potentials (\ref{grain}) 
and (\ref{graout}) into Eq.~(\ref{F0full}) to derive 
the leading-order solution to $\phi(r)$.
In doing so, second-order gravitational potentials like 
$2\phi \Psi'^2$ are neglected relative to 
their first-order contribution.

Employing this prescription for $r<r_*$, 
Eq.~(\ref{F0full}) reduces to 
\be
\frac{d}{dr} \left( r^2 \phi' \right)
+\phi_0 \Phi_* \left( 1-2 \beta_4 \phi_0^2 
\right) \frac{r^2}{r_*^2} \simeq 0\,.
\label{dreq}
\ee
Under the boundary condition $\phi'(0)=0$, the 
integrated solution to Eq.~(\ref{dreq}) is given by 
\be
\phi'(r) \simeq -\dfrac{\phi_0 \Phi_* (1-2\beta_4\phi_0^2)}
{3r_*^2}r\,.
\label{phidr}
\ee
We then obtain $\phi(r)$ in the form (\ref{phifr}) with 
$f(r)=-\phi_0 \Phi_* (1-2\beta_4 \phi_0^2)r^2/(6r_*^2)$. 
Provided that the term $1-2\beta_4 \phi_0^2$ is at most 
of the order of 1, the condition $|f(r)| \ll |\phi_0|$ is 
well satisfied.

For $r>r_*$, substituting Eq.~(\ref{graout}) into Eq.~(\ref{F0full}) 
and picking up first-order contributions of $\Phi$ and $\Psi$, 
it follows that 
\be
\frac{d}{dr} \left( r^2 \phi' \right) \simeq 0\,.
\label{phidr0}
\ee
The integrated solution to Eq.~(\ref{phidr0}) is given by 
$\phi'(r)=C/r^2$, where the constant $C$ is known by 
matching solutions at $r=r_*$.
Then, the resulting solution for the radius $r>r_*$ reads
\be
\phi' (r) \simeq -\frac{\phi_0 \Phi_*(1-2\beta_4\phi_0^2)}{3r^2}r_* \,.
\label{phiout}
\ee
Unlike Ref.~\cite{screening}, we have taken into account the term 
$2\beta_4 \phi_0^2$ without necessarily assuming the condition 
$|\beta_4| \phi_0^2 \ll 1$.

To derive corrections to the gravitational potentials (\ref{graout}) 
induced by the temporal vector component $\phi$ outside the 
compact body, we substitute Eqs.~(\ref{chi0}) and (\ref{phiout}) into 
Eqs.~(\ref{eq00}) and (\ref{eq11}) under the approximation 
of weak gravity. Then, for $r>r_*$, the gravitational potentials 
approximately obey 
\ba
&&
\frac{2M_{\rm pl}^2}{r}\Phi'+\frac{2M_{\rm pl}^2}{r^2} \Phi 
\simeq -\frac{\phi_0^2 \Phi_*^2 r_*^2}
{18r^4} (1-4\beta_4^2 \phi_0^4)\,,\\
&&
\frac{2M_{\rm pl}^2}{r}\Psi'-\frac{2M_{\rm pl}^2}{r^2} \Phi 
\simeq \frac{\phi_0^2 \Phi_*r_*}{18r^4} 
\left[ 12\beta_4 \phi_0^2 r (1-4\beta_4 \phi_0^2)+\Phi_*r_* 
\right]\,, 
\ea
which are integrated to give 
\ba
& &
\Phi(r) \simeq \frac{\Phi_* r_*}{6r} \left[ 
1+\frac{\phi_0^2 \Phi_* r_*}{6M_{\rm pl}^2 r}(1-4\beta_4^2 \phi_0^4)
\right]\,,
\label{Phiap1}\\
& &
\Psi(r) \simeq  -\frac{\Phi_* r_*}{6r} \left[
1+\frac{2\beta_4 \phi_0^4}{M_{\rm pl}^2}(1-4\beta_4 \phi_0^2)
+\frac{\phi_0^2 \Phi_* r_*}{6M_{\rm pl}^2 r} \right]\,.
\label{Psiap1}
\ea
As long as the correction terms in the square brackets of 
Eqs.~(\ref{Phiap1}) and (\ref{Psiap1}) are much smaller 
than 1, the post-Newtonian parameter $\gamma=-\Phi/\Psi$ 
reduces to 
\be
\gamma \simeq 1-\frac{2\beta_4 \phi_0^4}
{M_{\rm pl}^2}\left( 1-4\beta_4\phi_0^2 \right)\,.
\ee
On using the local gravity constraint 
$|\gamma-1|<2.3 \times 10^{-5}$ \cite{Will}, 
we obtain the bound 
\be
\left| \beta_4 \phi_0^4 (1-4\beta_4\phi_0^2) \right|
<1 \times 10^{-5}M_{\rm pl}^2\,.
\label{con1}
\ee
If $|\beta_4|\phi_0^2 \ll 1$, then the condition (\ref{con1}) 
translates to $|\beta_4|\phi_0^4<1 \times 10^{-5}M_{\rm pl}^2$. 
For $\phi_0$ of the order of $M_{\rm pl}$ the latter condition
gives the bound $|\beta_4| \lesssim 10^{-5}/M_{\rm pl}^2$, 
in which case the condition $|\beta_4|\phi_0^2 \ll 1$ is 
automatically satisfied. 
For the theories with $\beta_3=0$ the perfect screening 
of the longitudinal mode occurs even with the new term ${\cal L}_4^{\rm N}$, 
so only the temporal vector component $\phi$ 
gives rise to corrections to gravitational potentials.

\subsection{$\beta_3\neq 0$}

In the presence of the cubic interaction $G_3(X)$, 
$\chi'$ does not generally vanish. If the effect of the coupling $\beta_3$ 
always dominates over those of $\beta_4$ and $d_4$, 
then the resulting field profiles of $\phi$ and $\chi$ 
are practically the same as those derived in Ref.~\cite{screening} 
for $\beta_4=0=d_4$. 
Since we are interested in how the new interaction $f_4(X)=d_4X^n$ 
affects the screening mechanism, we focus on the case in which 
effects of the couplings $d_4$ and $\beta_4$ dominate 
over that of $\beta_3$ at least for the distance relevant 
to the solar system.

We employ the weak-gravity approximation 
and assume the condition 
\be
\chi'^2 \ll \phi^2\,,
\label{chias}
\ee
which can be justified after deriving the solution to $\chi'$. 
Then, Eqs.~(\ref{F0full}) and (\ref{F1full}) reduce, 
respectively, to 
\ba
& &
\frac{d}{dr} \left( r^2 \phi' \right)-\beta_3 \phi \frac{d}{dr} 
\left( r^2 \chi' \right)-4\beta_4\phi \frac{d}{dr} 
\left( r \chi'^2 \right)+2\phi \frac{d}{dr} 
\left( r^2 \Psi' \right)-4\beta_4 \phi^3 \left( \Phi+r\Phi' \right) 
\nonumber \\
& & 
-2^{2-n} d_4 \phi^{2n+1} \chi' \left[ (2+n) (2r\chi''+\chi')
+r\chi' \left\{ (3+2n)\Psi'-(5+2n)\Phi' \right\} \right]
\simeq 0\,,
\label{dyeq1} \\
& &
\chi' \simeq -\frac{\beta_3 r (r\phi\phi'+2\chi'^2+r\phi^2 \Psi')}
{4\beta_4[2r\phi \phi'+\chi'^2+\phi^2 (\Phi+r\Psi')]+
2^{2-n}d_4\phi^{2n}[(2+n)(2r\phi \phi'+\chi'^2)+r\phi^2
(2n\Psi'+3\Psi'-\Phi')]}\,.
\label{dyeq2}
\ea
In what follows, we derive the solutions to Eqs.~(\ref{dyeq1}) 
and (\ref{dyeq2}) for several different radii.
As before, we search for solutions of the temporal component
in the form (\ref{phifr}).

\subsubsection{$r<r_*$}

For the distance $r<r_*$, the leading-order solution to $\phi'(r)$ 
can be derived by taking the limit $\chi' \to 0$ in Eq.~(\ref{dyeq1}). 
This is equivalent to Eq.~(\ref{phidr}), i.e., 
\be
\phi'(r) \simeq -\dfrac{\rho_0 \phi_0 (1-2\beta_4\phi_0^2)}
{3M_{\rm pl}^2}r\,.
\label{phidr2}
\ee
Provided that the condition 
\be
\chi'^2 \ll r |\phi \phi'|
\label{chiphi}
\ee
is satisfied, substitutions of Eqs.~(\ref{grain}) and 
(\ref{phidr2}) into Eq.~(\ref{dyeq2}) lead to
\be
\chi'(r) \simeq -\frac{\beta_3 (1-4\beta_4 \phi_0^2)}
{8\beta_4 (1-4\beta_4 \phi_0^2)+2^{2-n}d_4\phi_0^{2n}
[7+2n-8(2+n)\beta_4 \phi_0^2]}r\,.
\label{chiin}
\ee
For $|\beta_4|\phi_0^2 \ll 1$, the condition (\ref{chiphi})
translates to 
\be
\varepsilon \equiv 
\frac{3\beta_3^2 M_{\rm pl}^2}
{[8\beta_4+2^{2-n}d_4(7+2n)\phi_0^{2n}]^2 \rho_0\phi_0^2}
\ll 1\,.
\label{varcon}
\ee
Since the solution (\ref{phidr2}) satisfies the inequality $|r\phi'| \ll |\phi|$, 
the assumption (\ref{chias}) is justified under the condition 
(\ref{chiphi}). 

Substituting Eq.~(\ref{chiin}) into Eq.~(\ref{dyeq1}) 
with $|\beta_4|\phi_0^2 \ll 1$, the next-to-leading 
order solution to $\phi'(r)$ is given by 
\be
\phi'(r) \simeq -\frac{\rho_0 \phi_0}{3M_{\rm pl}^2}
(1+\delta_1) r\,,\qquad 
\delta_1 \equiv \frac{3 \cdot 2^{n-2} \beta_3^2 M_{\rm pl}^2
[2^n \beta_4+(5+n)d_4\phi_0^{2n}]}
{[2^{n+1}\beta_4+(7+2n)d_4\phi_0^{2n}]^2 \rho_0}\,.
\label{delta}
\ee
Provided that 
{\bf 
\be
2^{-n}|d_4| \phi_0^{2n+2} \lesssim 1\,, 
\label{cond4}
\ee
}
$|\delta_1|$ is at most of the order of $\varepsilon$.
This means that, under the condition (\ref{varcon}), 
the correction $\delta_1$ to the leading-order solution (\ref{phidr2}) 
can be neglected. 
The above results show that, for $r<r_*$, both $|\chi'(r)|$ 
and $|\phi'(r)|$ linearly grow in $r$.

\subsubsection{$r_*<r<r_t$}

For $r>r_*$ the leading-order gravitational potentials are given by 
Eq.~(\ref{graout}), so this causes the decrease of $|\phi'(r)|$ as 
in Eq.~(\ref{phiout}).
Since there is a transition radius $r_t$ at which $\chi'^2$ grows 
to the same order as $r|\phi \phi'|$, we first study the behavior of 
solutions for the distance $r_*<r<r_t$. 
In this regime, we can neglect the terms $\chi'^2$ appearing on 
the r.h.s. of Eq.~(\ref{dyeq2}). Plugging Eq.~(\ref{graout}) into 
Eq.~(\ref{dyeq2}), it follows that 
\be
\chi'(r) \simeq 
-\frac{\beta_3}{8[\beta_4+2^{-n}(2+n)d_4\phi_0^{2n}]}r\,.
\label{chiou}
\ee

Substituting this into Eq.~(\ref{dyeq1}), we obtain 
the integrated solution 
\be
r^2 \phi'(r)+\frac{2^{n-4}\beta_3^2 \phi_0r^3}
{2^n \beta_4+(2+n)d_4 \phi_0^{2n}}={\cal C}\,.
\label{phidr3}
\ee
The constant ${\cal C}$ is fixed by matching 
Eq.~(\ref{phidr3}) with Eq.~(\ref{phidr2}) at $r=r_*$.
The resulting solution to $\phi'(r)$ for the distance 
$r_*<r<r_t$ is given by 
\be
\phi'(r) \simeq -\frac{\rho_0 \phi_0 r_*^3}
{3M_{\rm pl}^2 r^2} 
\left[ 1-2\beta_4 \phi_0^2 +\delta_2 
\left( \frac{r^3}{r_*^3} -1 \right) \right]\,,\qquad 
\delta_2 \equiv \frac{3 \cdot 2^{n-4}\beta_3^2 M_{\rm pl}^2}
{[2^n \beta_4+(2+n)d_4\phi_0^{2n}]\rho_0}\,.
\label{phiou}
\ee
The quantity $|\delta_2|$ is as small as $|\delta_1|$ ($\lesssim \varepsilon \ll 1$).
Provided that $|\delta_2| r^3/r_*^3 \ll 1$, $|\phi'(r)|$ decreases 
in proportion to $r^{-2}$.

The growth of $|\chi'(r)|$ saturates for the distance
$r_t$ at which the numerator of Eq.~(\ref{dyeq2}) is close to 0. 
Employing the solutions (\ref{chiou}) and (\ref{phiou}) and neglecting the 
term $\delta_2(r^3/r_*^3-1)$, the transition radius can be estimated as
\be
r_t=r_* \left\{ \frac{16(1-4\beta_4\phi_0^2)}{\varepsilon}
\left[ \frac{\beta_4+2^{-n}(2+n)d_4\phi_0^{2n}}
{8\beta_4+2^{2-n}(7+2n)d_4\phi_0^{2n}} \right]^2 \right\}^{1/3}\,.
\ee
For $|\beta_4|\phi_0^2 \ll 1$ we have $r_t \approx r_*/\epsilon^{1/3}$, 
so $r_t$ is larger than $r_*$ for $\varepsilon \ll 1$.

\subsubsection{$r>r_t$}

For the distance $r>r_t$, the growth of $|\chi'(r)|$ saturates in such a way 
that the numerator on the r.h.s. of Eq.~(\ref{dyeq2}) is close to 0, i.e., 
\be
r\phi \phi'+2\chi'^2+\frac{\rho_0 \phi^2r_*^3}
{6M_{\rm pl}^2 r} \simeq 0\,.
\label{chiex}
\ee
In this case, the terms associated with the coupling $\beta_3$ 
dominate over those related to $\beta_4$ and $d_4$ in 
Eq.~(\ref{dyeq2}), e.g., $|\beta_3 r| \gg |\beta_4 \chi'|$ and 
$|\beta_3 r| \gg |d_4 \phi^{2n} \chi'|$. 
Then, the terms containing $\beta_4$ and $d_4$ in Eq.~(\ref{dyeq1}) 
can be neglected relative to the $\beta_3$-dependent term. 
This leads to the integrated solution
\be
r^2 \phi' -\beta_3 \phi_0 r^2 \chi' \simeq 
-\frac{\rho_0 \phi_0r_*^3}{3M_{\rm pl}^2} (1-2\beta_4 \phi_0^2)\,.
\label{phisort}
\ee
The r.h.s. of Eq.~(\ref{phisort}) corresponds to the integration 
constant determined by matching the solutions at $r=r_t$.
We can explicitly solve Eqs.~(\ref{chiex}) and (\ref{phisort}) 
for $\phi'(r)$ and $\chi'(r)$, respectively, as 
\ba
\phi'(r) & \simeq&-\frac{\rho_0 \phi_0 r_*^3}
{3M_{\rm pl}^2 r^2}{\cal G}(\eta)\,,\label{phiex}\\
\chi'(r) &\simeq& \pm \sqrt{\frac{\rho_0 \phi_0^2 r_*^3}
{6M_{\rm pl}^2 r} \left[ {\cal G}(\eta)-\frac12 
\right]}\,,\label{chiex2}
\ea
where 
\ba
\eta &\equiv& \frac{3\beta_3^2 \phi_0^2M_{\rm pl}^2}
{4\rho_0} \frac{r^3}{r_*^3}
=4\phi_0^4 (1-4\beta_4 \phi_0^2 ) 
\left[ \beta_4+2^{-n}(2+n)d_4 \phi_0^{2n} \right]^2 
\frac{r^3}{r_t^3}
\label{eta}
\,,\\
{\cal G}(\eta) &\equiv& (1+\eta-2\beta_4 \phi_0^2) 
\left[ 1-\sqrt{1-\frac{(1-2\beta_4 \phi_0^2)^2+\eta}
{(1+\eta-2\beta_4 \phi_0^2)^2}} \right]\,.
\ea
Provided that 
\be
\delta \equiv 4\phi_0^4 (1-4\beta_4 \phi_0^2 ) 
\left[ \beta_4+2^{-n}(2+n)d_4 \phi_0^{2n} \right]^2 \ll1\,,
\ee
the behavior of solutions changes at the distance $r_v$ 
satisfying $\eta=1$, i.e., 
\be
r_v=\frac{r_t}{\delta^{1/3}}\,.
\ee
{}From the first equality of Eq.~(\ref{eta}) the distance 
$r_v$ itself does not depend on $\beta_4$ and $d_4$, but the ratio $r_v/r_t$ is dependent on these couplings.

For $r \ll r_v$ the field profiles read
\be
\phi'(r) \simeq -\frac{\rho_0 \phi_0 r_*^3}
{3M_{\rm pl}^2 r^2} (1-2\beta_4 \phi_0^2)\,,\qquad
\chi'(r) \simeq \pm \sqrt{\frac{\rho_0 \phi_0^2 r_*^3}
{12M_{\rm pl}^2 r} (1-4\beta_4 \phi_0^2)}\,,
\label{phichi}
\ee
whereas, for $r \gg r_v$, we have
\be
\phi'(r) \simeq -\frac{\rho_0 \phi_0 r_*^3}
{6M_{\rm pl}^2 r^2}\,,\qquad
\chi'(r) \simeq \pm \frac{\rho_0 r_*^3}
{6\beta_3 M_{\rm pl}^2r^2}\,,
\ee
both of which do not contain the coupling $d_4$.

\subsubsection{Gravitational potentials for $r>r_*$}

The structure of the field profiles derived above 
is similar to that obtained in Ref.~\cite{screening} 
for $d_4=0$, apart from the difference of some 
coefficients of $\chi'(r)$. In the limit that $\beta_3 \to 0$, 
the values of $\chi'(r)$ given in Eqs.~(\ref{chiin}) 
and (\ref{chiou}) vanish with $r_t \approx r_*/\varepsilon^{1/3} \to \infty$.
Hence, for small $\beta_3$,  the effect of the new 
coupling $d_4$ on the longitudinal scalar 
is unimportant.

The equations of motion (\ref{eq00}) and (\ref{eq11}) 
of gravitational potentials contain $f_4$-dependent 
terms. For $r>r_*$ we shall estimate such contributions to the 
leading-order gravitational potentials (\ref{graout}).
We also assume that the term $|\beta_4| \phi_0^2$
is much smaller than 1.

In the regime $r_*<r<r_t$, we substitute the solutions (\ref{chiou}) and (\ref{phiou}) into Eqs.~(\ref{eq00}) and (\ref{eq11}) by neglecting the $\delta_2 (r^3/r_*^3-1)$ term and integrate them under the condition (\ref{chias}). 
This leads to the following solutions 
\ba
\Phi (r)
&\simeq& \frac{\Phi_* r_*}{6r}
\left[ 1+\frac{\phi_0^2\Phi_*}{6M_{\rm pl}^2 x}
-\frac{2^{n+3}\beta_4 \phi_0^4 \epsilon_0 x^3}{M_{\rm pl}^2}
-\frac{8d_4(2n+3)\phi_0^{2n+4}\epsilon_0 x^3}{M_{\rm pl}^2} \right]\,,\\
\Psi (r)
&\simeq&-\frac{\Phi_* r_*}{6r}
\left[ 1+\frac{\phi_0^2\Phi_*}{6M_{\rm pl}^2 x}
+\frac{2\beta_4 \phi_0^4 (1+2^{n+2}\epsilon_0 x^3)}{M_{\rm pl}^2}
+\frac{8d_4(n+3)\phi_0^{2n+4}\epsilon_0 x^3}{M_{\rm pl}^2}
\right]\,,
\ea
where $x \equiv r/r_*$ and 
\be
\epsilon_0 \equiv \frac{3\beta_3^2 M_{\rm pl}^2}
{2^{n+8} \left[\beta_4+2^{-n}d_4(2+n)\phi_0^{2n}\right]^2 \rho_0\phi_0^2}\,. 
\ee
In deriving the above solutions, we have neglected the 
term $\Phi_* \epsilon_0 x^2$ relative to 1.
The post-Newtonian parameter $\gamma=-\Phi/\Psi$ reduces to 
\be
\gamma \simeq 1-\frac{2\beta_4 \phi_0^4 (2^{n+3}\epsilon_0 x^3+1)}
{M_{\rm pl}^2} -\frac{24d_4(n+2)\phi_0^{2n+4}\epsilon_0x^3}{M_{\rm pl}^2}\,. 
\label{gamma1}
\ee

For $n=-2$, which corresponds to the case in which 
solid angle deficit singularities appear in the scalar limit 
$A^{\mu} \to \nabla^{\mu} \chi$ 
(i.e., GLPV theories), the effect of the coupling $d_4$ appears 
in $\Phi(r)$ and $\Psi(r)$, but it disappears in $\gamma$.
Since the term $2\epsilon_0 x^3$ is at most of the order 
of 1 at $r=r_t$, the solar-system bound on 
$\beta_4 \phi_0^4$ is similar to Eq.~(\ref{con1}), 
i.e., $|\beta_4| \phi_0^4 \lesssim 10^{-5}M_{\rm pl}^2$. 

For $n\neq-2$, the contribution of the coupling $d_4$ survives
in $\gamma$. At $r=r_t$, Eq.~(\ref{gamma1}) reduces to 
\be
\gamma \simeq 1-\frac{3\beta_4\phi_0^4}{M_{\rm pl}^2}
-\frac{3\cdot 2^{-n-1}(n+2)d_4\phi_0^{2n+4}}{M_{\rm pl}^2}\,.
\label{gamma2}
\ee
The second term on the r.h.s. of Eq.~(\ref{gamma2}) is similar 
to that for $n=-2$, so we obtain 
the bound $|\beta_4| \phi_0^4 \lesssim 10^{-5}M_{\rm pl}^2$. 
Under the condition (\ref{cond4}), the third term on the r.h.s. of 
Eq.~(\ref{gamma2}) is compatible with the solar-system 
constraint for $\phi_0\lesssim10^{-3}M_{\rm pl}$. 

At the distance $r_t<r<r_v$, we substitute the solutions (\ref{phichi}) 
into Eqs.~(\ref{eq00}) and (\ref{eq11}). 
Then, we obtain the integrated solution 
\ba
&&\hspace{-.4cm}
\Phi (r)
\simeq 
\frac{\Phi_* r_*}{6r}
\left[ 1+\frac{\phi_0^2\Phi_*}{6M_{\rm pl}^2x}
-\frac{2^{n/2+3}\beta_4\phi_0^4\sqrt{\epsilon_0}x^{3/2}}{M_{\rm pl}^2}
-\frac{2^{-n}\phi_0^{2n+4}d_4}{M_{\rm pl}^2}
\left\{2^{n/2+3}(n+2)\sqrt{\epsilon_0}x^{3/2}+\frac{(10n+31)\Phi_*}{24x}\right\}
\right]\,,\label{Phil}\\
&&\hspace{-.4cm}
\Psi (r)
\simeq -\frac{\Phi_* r_*}{6r}
\left[ 1+\frac{\phi_0^2\Phi_*}{6 M_{\rm pl}^2x}
+\frac{\beta_4\phi_0^4(2^{n/2+5}\sqrt{\epsilon_0}x^{3/2}+3)}{2M_{\rm pl}^2}
+\frac{2^{-n}\phi_0^{2n+4}d_4}{M_{\rm pl}^2}
\left\{2^{n/2+4}(n+2)\sqrt{\epsilon_0}x^{3/2}-\frac{(8n+27)\Phi_*}{24x}\right\}
\right]\,.\notag\\
\label{Psil}
\ea
The post-Newtonian parameter is given by 
\be
\gamma\simeq1
-\frac{3\beta_4\phi_0^4(2^{n/2+4}\sqrt{\epsilon_0}x^{3/2}+1)}{2M_{\rm pl}^2} 
-\frac{2^{-n-2}(n+2)\phi_0^{2n+4}d_4}{3M_{\rm pl}^2}
\left(9\cdot2^{n/2+5}\sqrt{\epsilon_0}x^{3/2}+\frac{\Phi_*}{x}\right)\,. 
\label{gamma3}
\ee

For $n=-2$, even though there are terms containing $d_4$ in the gravitational 
potentials (\ref{Phil}) and (\ref{Psil}), 
the effect of the coupling $d_4$ on $\gamma$ disappears.
As in the case $d_4=0$ studied in Ref.~\cite{screening}, the local gravity 
constraint is satisfied for $\phi_0 \lesssim 10^{-3}M_{\rm pl}$.

For $n\neq-2$, the contribution from the coupling $d_4$ remains in $\gamma$. 
At $r=r_v$, the dimensionless distance $x_v=r_v/r_*$ satisfies
$\sqrt{\epsilon_0}x_v^{3/2}=[2^{n/2+3}\phi_0^2\{\beta_4
+2^{-n}d_4(n+2)\phi_0^{2n}\}]^{-1}$. 
Substituting this relation into Eq.~(\ref{gamma3}), ignoring the last term 
$\Phi_*/x$ in Eq.~(\ref{gamma3}), 
and taking the limit that the coupling $d_4$ dominates over $\beta_4$ such that 
$|2^{-n}d_4(n+2)\phi_0^{2n}| \gg |\beta_4|$, the post-Newtonian parameter simply 
reduces to $\gamma\simeq1-3\phi_0^2/M_{\rm pl}^2$.
Hence the resulting bound on $\phi_0$ is the same as that for $n=-2$. 

\section{Conclusions}
\label{consec}

The beyond-generalized Proca theories were constructed 
in such a way that they recover the shift-symmetric GLPV theories 
in the scalar limit. In GLPV theories, solid angle deficit singularities 
arise in the case where the deviation from Horndeski theories 
(weighed by the parameter $\alpha_{\rm H}$) does not 
vanish at the center of a spherically symmetric body ($r=0$). 
The appearance of solid angle deficit singularities is associated 
with the fact that the gravitational potential $\Phi(r)$ 
contains the non-vanishing constant 
$-(1/2)\ln (1+\alpha_{\rm H})$ in the limit $r \to 0$, 
see Eq.~(\ref{Phiso}).

In this paper, we derived spherically symmetric solutions 
around the center of the compact body in quartic-order 
beyond-generalized Proca theories. For the model described by the functions (\ref{model2}) the resulting gravitational potentials around $r=0$ are given by Eqs.~(\ref{sol1})-(\ref{sol2}), so they satisfy the conditions (\ref{Phire})-(\ref{Psire}) 
for the absence of solid angle deficit singularities. 
In fact, we showed that solid angle deficit singularities are generally 
absent in quartic-order beyond-generalized Proca theories 
with arbitrary functions $G_{2,3,4},g_4,f_4$.
This is mainly attributed to the fact that Eq.~(\ref{Phiap}) 
contains the term $2\Phi'/r-(1-e^{2\Phi})/r^2$, 
which demands the two conditions (\ref{Phi01}) for 
the consistency of Eq.~(\ref{Phiap}). 
Existence of the temporal component $\phi$ in 
beyond-generalized Proca theories leads to 
the vanishing constant $\Phi_0$ in the expansion 
of $\Phi$ around $r=0$.
This is not the case for GLPV theories 
in which the extra term $1+\alpha_{\rm H}$ 
in the square bracket of Eq.~(\ref{Phianaeq}) 
gives rise to the non-vanishing constant $\Phi_0$. 

We also studied the Vainshtein mechanism in the presence of the 
quartic-order beyond-generalized Proca Lagrangian with 
the coupling $f_4(X)=d_4X^n$ besides vector-Galileon terms. 
If the cubic vector-Galileon term is absent ($\beta_3=0$),  
we obtained the solution where the derivative $\chi'$ of the 
longitudinal scalar exactly vanishes. 
In this case, the beyond-generalized Proca interaction 
does not give rise to any contribution to the temporal 
vector component $\phi$, whose solution 
outside the body ($r>r_*$) is given by Eq.~(\ref{phiout}).  
Provided that the condition (\ref{con1}) is satisfied, the corrections
to the gravitational potentials $\Phi$ and $\Psi$ induced by 
the vector field are sufficiently small such that the theories 
are compatible with solar-system constraints. 

We also derived the vector-field profiles in the case where 
the coupling $\beta_3$ is present besides $\beta_4$ and $d_4$.
The radial dependence of the vector field is similar to that for $d_4=0$ 
apart from the difference of some coefficients, so the Vainshtein 
mechanism works in a similar way to that discussed 
in Ref.~\cite{screening}. We also found that there are corrections 
to the leading-order gravitational potentials $\Phi$ and $\Psi$ from 
the new coupling $d_4$, but as long as the coupling $d_4$ is in the 
range with $\phi_0 \lesssim 10^{-3}M_{\rm pl}$, the solar-system 
bound on $\gamma=-\Phi/\Psi$ is well satisfied.

We have thus shown that the quartic-order beyond-generalized 
Proca theories have a nice feature for avoiding the problem of 
solid angle deficit singularities at the center of the compact body, while 
allowing the success of the Vainshtein screening outside 
the body. Since the appearance of solid angle deficit singularities in GLPV 
theories is intrinsically related to the beyond-Horndeski Lagrangian
at quartic order \cite{conical2}, we anticipate that the absence of 
solid angle deficit singularities beyond-generalized Proca theories should persist 
in more general cases containing the fifth- and sixth-order Lagrangians 
derived in Ref.~\cite{HKT}.
Nevertheless this requires further detailed study, so we leave the analysis 
for the derivation of spherically symmetric solutions 
in full beyond-generalized Proca theories as a future work.

\section*{Acknowledgements}

LH thanks financial support from Dr.~Max R\"ossler, 
the Walter Haefner Foundation and the ETH Zurich
Foundation. RK is supported by the Grant-in-Aid for Research Activity
Start-up of the JSPS No.\,15H06635. 
ST is supported by the Grant-in-Aid for Scientific Research Fund of 
the JSPS No.~16K05359 and MEXT KAKENHI Grant-in-Aid for 
Scientific Research on Innovative Areas ``Cosmic Acceleration'' 
(No.\,15H05890).

\appendix

\section{Coefficients}

The coefficients in Eqs.~(\ref{eq00})-(\ref{eq1})
are given by 
\ba
&&
C_{1}=4X_{\phi} ( 1-2g_{4}-4X_{\phi}g_{4,X} ) 
\,,\quad
C_{2}=4X_{\phi}{\chi'}G_{3,X}+2\phi{\phi'}
e^{2\Psi} ( 1-2g_{4}-4X_{\phi}g_{4,X} ) 
\,,\notag\\
&&
C_{3}=-32X_{\phi}X_{\chi} \left[ G_{4,{XX}}+3f_{4}+2
 ( X_{\phi}+X_{\chi} ) f_{4,X} \right] 
\,,\quad
C_{4}=-2G_{3,X} ( X_{\phi}+X_{\chi} ) {\chi'}
\,,\notag\\
&&
C_{5}=-4G_{4}+8 ( X_{\phi}+2X_{\chi} ) G_{4,
X}+16X_{\chi} ( X_{\phi}+X_{\chi} ) G_{4,{XX}
}+16X_{\chi} ( 7X_{\phi}+5X_{\chi} ) f_{4}+
32X_{\chi} ( X_{\phi}+X_{\chi} ) ^{2}f_{4,X}
\,,\notag\\
&&
C_{6}=-e^{2\Phi} ( G_{2}-2X_{\phi}G_{2,X}) 
+ \left[ {\chi'}\phi{\phi'}e^{2\Psi}+2
 ( X_{\phi}+X_{\chi} ) {\chi''} \right] G_{3,X}+
e^{2\Psi}{\phi'}^{2} ( 1-2g_{4}-4X_{\phi}g_{4,X} )/2
\,,\notag\\
&&
C_{7}=4X_{\phi}{\chi'}G_{3,X}+4 \left[ G_{4,X}+2
 ( X_{\phi}+X_{\chi} ) G_{4,{XX}}+2 ( 
 5X_{\phi}+4X_{\chi} ) f_{4}+4 ( X_{\phi}+X_{\chi} ) ^{2}
f_{4,X} \right] e^{-2\Phi}{\chi'}{
\chi''}\notag\\
&&\hspace{1cm}
-8X_{\chi} \left[ G_{4,{XX}}+3f_{4}+2 ( X_
{\phi}+X_{\chi} ) f_{4,X} \right] e^{2\Psi}\phi
{\phi'}
\,,\notag\\
&&
C_{8}=2 ( 1-e^{2\Phi} ) \left( G_{4}
-2X_{\phi}G_{4,X} \right)-4X_{\chi} G_{4,X}
-8X_{\phi}X_{\chi}G_{4,{XX}}
\notag\\
&&
\hspace{1cm}
-8X_{\chi} ( 5X_{\phi}+X_{\chi} ) f_{4}-16X_{\phi}X_{\chi} ( 
X_{\phi}+X_{\chi} ) f_{4,X}
\,,\notag\\
&&
C_{9}=4X_{\phi} ( 1-2g_{4}-4X_{\chi}g_{4,X} ) 
\,,\quad
C_{10}=-2G_{3,X} ( X_{\phi}-X_{\chi} ) {\chi'}
+2\phi{\phi'}e^{2\Psi} ( 1-2g_{4}-4X_{\chi}g_{4,X} ) 
\,,\notag\\
&&
C_{11}=4G_{4}+8 ( X_{\phi}-2X_{\chi} ) G_{4,
X}+16X_{\chi} ( X_{\phi}-X_{\chi} ) G_{4,{XX
}}+16X_{\chi} ( 3X_{\phi}-5X_{\chi} ) f_{4}
+32X_{\chi} ( X_{\phi}^{2}-X_{\chi}^{2} ) f_{4
,X}
\,,\notag\\
&&
C_{12}=-e^{2\Phi} ( G_{2}-2G_{2,X}X_{\chi}
 ) -e^{2\Psi}G_{3,X}{\chi'}\phi{\phi'}+
{\phi'}^{2}e^{2\Psi} ( 1-2g_{4}-4X_{\chi}g_{4,X} )/2
\,,\notag\\
&&
C_{13}=4G_{3,X}{\chi'}X_{\chi}+8e^{-2\Phi}X_
{\phi}{\chi'}{\chi''}f_{4}+4 \left[ 2X_{\chi}G_{4,
{XX}}+10f_{4}X_{\chi}+G_{4,X}+4X_{\chi} ( X_{
\phi}+X_{\chi} ) f_{4,X} \right] e^{2\Psi}\phi{
\phi'}
\,,\notag\\
&&
C_{14}=2 ( 1-e^{2\Phi} ) G_{4}-4 ( 2
-e^{2\Phi} ) X_{\chi}G_{4,X}-8G_{4,{XX}}{X
_{\chi}}^{2}-8X_{\chi} ( X_{\phi}+5X_{\chi} ) f
_{4}-16{X_{\chi}}^{2} ( X_{\phi}+X_{\chi} ) f_{4
,X}
\,,\notag\\
&&
C_{15}=2G_{4}-8f_{4}{X_{\chi}}^{2}+4 ( X_{\phi}-X
_{\chi} ) G_{4,X}
\,,\quad
C_{16}=8f_{4}X_{\phi}X_{\chi}
\,,\notag\\
&&
C_{17}=-4X_{\phi} ( 1-2g_{4} ) +2G_{4}+4
 ( 4X_{\phi}-X_{\chi} ) G_{4,X}+8X_{\phi}
 ( X_{\phi}-X_{\chi} ) G_{4,{XX}}-8{X_{\chi}
}^{2} ( f_{4}+2X_{\phi}f_{4,X} ) 
\,,\notag\\
&&
C_{18}=-2G_{4}+8X_{\chi} ( 3X_{\phi}+5X_{\chi}
 ) f_{4}+16X_{\chi} ( X_{\phi}^{2}+X_{\chi}^{2} ) f_{4,X}
 -4 ( X_{\phi}-2X_{\chi} ) G_{4
,X}-8X_{\chi} ( X_{\phi}-X_{\chi} ) G_{4,{XX
}}
\,,\notag\\
&&
C_{19}=-8X_{\chi}X_{\phi} ( 3f_{4}+2X_{\chi}f_{4,X} ) 
\,,\notag\\
&&
C_{20}=2X_{\phi}{\chi'}G_{3,X}-2\phi e^{2\Psi
} ( 1-2g_{4} ) {\phi'}
\notag\\
&&
\hspace{1cm}
+4 \left[ 3G_{4,X}+
 ( 2X_{\phi}-X_{\chi} ) G_{4,{XX}}+3f_{4}X_
{\chi}+2X_{\chi} ( X_{\phi}-X_{\chi} ) f_{4,X}
 \right] \phi{\phi'}e^{2\Psi}
 \notag\\
 &&\hspace{1cm}
 +2 \left[ G_{4,X}-2
 ( X_{\phi}-X_{\chi} ) G_{4,{XX}}+2 ( 3X_{\phi}+4X_{\chi} ) f_{4}
 +4 ( X_{\phi}^{2}+X_{\chi}^{2} ) f_{4,X} \right] e^{-2\Phi}{
\chi'}{\chi''}
\,,\notag\\
&&
C_{21}=2G_{4}+4 ( X_{\phi}-X_{\chi} ) G_{4,X}
-8X_{\phi}X_{\chi}G_{4,{XX}}-8X_{\chi} ( 3X_{\phi}+X_{\chi} ) f_{4}
-16X_{\phi}X_{\chi} ( X_{\phi}+X_{\chi} ) f_{4,X}
\,,\notag\\
&&
C_{22}=-2G_{3,X}{\chi'}X_{\chi}-2 \left[ G_{4,X}+2X_{\chi}G
_{4,{XX}}+2f_{4}X_{\chi}-4X_{\chi} ( X
_{\phi}-X_{\chi} ) f_{4,X} \right] \phi{\phi'}
e^{2\Psi}
\notag\\
&&\hspace{1cm}
-4e^{-2\Phi}X_{\phi} ( 5f_{4}+4X_{\chi}f_{4,X} ) {\chi'}{\chi''}
\,,\notag\\
&&
C_{23}=-2G_{4}+8X_{\chi} ( G_{4,X}+X_{\chi}G_{4,{XX}} ) 
+8X_{\chi} ( 3X_{\phi}+5X_{\chi}
 ) f_{4}+16{X_{\chi}}^{2} ( X_{\phi}+X_{\chi}
 ) f_{4,X}
\,,\notag\\
&&
C_{24}= \left[ G_{3,X}{\chi'}\phi{\phi'}+2 ( G_{4,X}+2f_{4}X_{\chi} ) \phi{\phi''}
+2 ( G_{4,X}+2G_{4,{XX}}X_{\phi}+2X_{\chi}f_{4}+4X_{\chi}X_{\phi}f_{4,X} ) {
\phi'}^{2} \right] e^{2\Psi}
\notag\\
&&\hspace{1cm}
- \left[ G_{2}+( 1-2g_{4} ) e^{2
\Psi-2\Phi}{\phi'}^{2}/2 \right] e^{2\Phi}
+2 \left[ X_{\chi}G_{3,X}- \left\{ G_{4,{XX}}-2 ( X_{\phi}-X_
{\chi} ) f_{4,X} \right\} e^{2\Psi-2\Phi}\phi{\phi'}{\chi'} \right] {\chi''}
\notag\\
&&\hspace{1cm}
+4X_{\phi}e^{-2\Phi} \left[  ( f_{4}+2X_{\chi}f_{4,X} ) {
\chi''}^{2}+f_{4}{\chi'}{\chi'''} \right] 
\,,\notag\\
&&
C_{25}=2 \left[ G_{4,X}-2X_{\chi}G_{4,{XX}}
-4X_{\chi}f_{4}-4X_{\chi} ( X_{\phi}+X_{\chi} ) f_{4,X} \right] 
\phi{\phi'}e^{2\Psi}
\notag\\
&&\hspace{1cm}
+2 \left[ G_{4,X}+
2X_{\chi}G_{4,{XX}}+4 ( X_{\phi}+2X_{\chi}
 ) f_{4}+4X_{\chi} ( X_{\phi}+X_{\chi} ) f_
{4,X} \right] e^{-2\Phi}{\chi'}{\chi''}\,,\notag\\
&&
D_{1}=-2 ( 1-2g_{4} ) \phi
\,,\quad
D_{2}=2\phi ( 1-2g_{4}-4X_{\chi}g_{4,X} ) 
\,,\notag\\
&&
D_{3}=G_{3,X}{\chi'}\phi-4e^{-2\Phi}\phi g_{4,X
}{\chi'}{\chi''}- ( 3-6g_{4}-4X_{\phi}g_{4,X}
 ) {\phi'}
\,,\notag\\
&&
D_{4}=-4 \left[ 1-2g_{4}+2X_{\chi}G_{4,{XX}}
+6X_{\chi}f_{4}+4X_{\chi} ( X_{\phi}+X_{\chi} ) f_{4
,X} \right] \phi
\,,\notag\\
&&
D_{5}= ( 1-2g_{4}-4X_{\chi}g_{4,X} ) {\phi'
}-G_{3,X}{\chi'}\phi
\,,\notag\\
&&
D_{6}=4 \left[ G_{4,X}+2X_{\chi}G_{4,{XX}}+10f_{4}X_{\chi}
+4X_{\chi} ( X_{\phi}+X_{\chi} ) f_{4,X}
 \right] \phi
\,,\notag\\
&&
D_{7}= ( G_{2,X}e^{2\Phi}+G_{3,X}{\chi''}+e^{2\Psi}g_{4,X}{\phi'}^{2} ) \phi
- ( 1-2g_{4} ) {\phi''}-2e^{-2\Phi}g_{4,X}{\phi'}{\chi'}{\chi''}
\,,\notag\\
&&
D_{8}=2 \left[ G_{3,X}{\chi'}+2 \left\{ G_{4,{XX}}+4
f_{4}+2 ( X_{\phi}+X_{\chi} ) f_{4,X} \right\} {
\chi'}{\chi''}e^{-2\Phi} \right] \phi-2 ( 1-2
g_{4} ) {\phi'}
\,,\notag\\
&&
D_{9}=2 \left[  ( e^{2\Phi}-1 ) G_{4,X}-2
X_{\chi}G_{4,{XX}}-8X_{\chi}f_{4}-4X_{\chi} ( X
_{\phi}+X_{\chi} ) f_{4,X} \right] \phi
\,,\notag\\
&&
D_{10}=8X_{\phi}g_{4,X}{\chi'}
\,,\quad
D_{11}=4g_{4,X}e^{2\Psi}\phi{\phi'}{\chi'}-2
 ( X_{\phi}-X_{\chi} ) G_{3,X} e^{2\Phi}
\,,\notag\\
&&
D_{12}=4 \left[ G_{4,X}-2 ( X_{\phi}-X_{\chi}
 ) G_{4,{XX}}-2 ( 3X_{\phi}-4X_{\chi}
 ) f_{4}-4 ( X_{\phi}^{2}-X_{\chi}^{2} ) 
f_{4,X} \right] {\chi'}
\,,\notag\\
&&
D_{13}=8{\chi'}X_{\phi}f_{4}
\,,\quad
D_{14}= ( g_{4,X}e^{2\Psi}{\phi'}^{2}-G_{2,X}
e^{2\Phi} ) {\chi'}-G_{3,X}e^{2\Psi+2
\Phi}{\phi'}\phi
\,,\notag\\
&&
D_{15}=4X_{\chi}G_{3,X}e^{2\Phi}-4 \left[ G_{4,{
XX}}+4f_{4}+2 ( X_{\phi}+X_{\chi} ) f_{4,X}
 \right] e^{2\Psi}\phi{\phi'}{\chi'}
\,,\notag\\
&&
D_{16}=2 \left[  ( 1-e^{2\Phi} ) G_{4,X}+2
X_{\chi}G_{4,{XX}}+8X_{\chi}f_{4}+4X_{\chi}
 ( X_{\phi}+X_{\chi} ) f_{4,X} \right] {\chi'}\,.
\ea
%



\begin{thebibliography}{99}

\bibitem{SNIa}
A.~G.~Riess \textit{et al.}
[Supernova Search Team Collaboration],
Astron.\ J.\  {\bf 116}, 1009 (1998) [astro-ph/9805201];
S.~Perlmutter \textit{et al.}
[Supernova Cosmology Project Collaboration],
Astrophys.\ J.\  {\bf 517}, 565 (1999) [astro-ph/9812133].

\bibitem{CMB}
D.~N.~Spergel {\it et al.} [WMAP Collaboration],
Astrophys.\ J.\ Suppl.\  {\bf 148}, 175 (2003)
[astro-ph/0302209];
P.~A.~R.~Ade {\it et al.} [Planck Collaboration],
Astron.\ Astrophys.\  {\bf 571}, A16 (2014)
[arXiv:1303.5076 [astro-ph.CO]].

\bibitem{BAO}
D.~J.~Eisenstein {\it et al.} [SDSS Collaboration],
Astrophys.\ J.\  {\bf 633}, 560 (2005)
[astro-ph/0501171].

\bibitem{Weinberg}
S.~Weinberg,
Rev.\ Mod.\ Phys.\  {\bf 61}, 1 (1989);
J.~Martin,
Comptes Rendus Physique {\bf 13}, 566 (2012)
[arXiv:1205.3365 [astro-ph.CO]];
A.~Padilla,
arXiv:1502.05296 [hep-th].

\bibitem{CST}
E.~J.~Copeland, M.~Sami and S.~Tsujikawa,
Int.\ J.\ Mod.\ Phys.\ D {\bf 15}, 1753 (2006)
[hep-th/0603057];
A.~Silvestri and M.~Trodden,
Rept.\ Prog.\ Phys.\  {\bf 72}, 096901 (2009)
[arXiv:0904.0024 [astro-ph.CO]];
S.~Tsujikawa,
arXiv:1004.1493 [astro-ph.CO];
T.~Clifton, P.~G.~Ferreira, A.~Padilla and C.~Skordis,
Phys.\ Rept.\  {\bf 513}, 1 (2012)
[arXiv:1106.2476 [astro-ph.CO]];
A.~Joyce, B.~Jain, J.~Khoury and M.~Trodden,
Phys.\ Rept.\  {\bf 568}, 1 (2015)
[arXiv:1407.0059 [astro-ph.CO]];
P.~Bull {\it et al.},
Phys.\ Dark Univ.\  {\bf 12}, 56 (2016)
[arXiv:1512.05356 [astro-ph.CO]];
L.~Amendola {\it et al.},
Living Rev.\ Rel.\  {\bf 16}, 6 (2013)
[arXiv:1206.1225 [astro-ph.CO]];
L.~Amendola {\it et al.},
arXiv:1606.00180 [astro-ph.CO].

\bibitem{quin} 
Y.~Fujii, Phys.\ Rev.\ D {\bf 26}, 2580 (1982);
C.~Wetterich, Nucl. \ Phys \ B. {\bf 302}, 668 (1988);
B.~Ratra and P.~J.~E.~Peebles,
Phys.\ Rev.\ D {\bf 37}, 3406 (1988);
T.~Chiba, N.~Sugiyama and T.~Nakamura,
Mon.\ Not.\ Roy.\ Astron.\ Soc.\  {\bf 289}, L5 (1997)
[astro-ph/9704199];
R.~R.~Caldwell, R.~Dave and P.~J.~Steinhardt,
Phys.\ Rev.\ Lett.\  {\bf 80}, 1582 (1998)
[astro-ph/9708069].

\bibitem{Brans} 
C.~Brans and R.~H.~Dicke,
Phys.\ Rev.\  {\bf 124}, 925 (1961).

\bibitem{Deffayet}
C.~Deffayet, G.~Esposito-Farese and A.~Vikman,
Phys.\ Rev.\  D {\bf 79}, 084003 (2009)
[arXiv:0901.1314 [hep-th]];
C.~Deffayet, S.~Deser and G.~Esposito-Farese,
Phys.\ Rev.\  D {\bf 80}, 064015 (2009)
[arXiv:0906.1967 [gr-qc]].

\bibitem{Gali2}
R.~Gannouji and M.~Sami,
Phys.\ Rev.\ D {\bf 82}, 024011 (2010)
[arXiv:1004.2808 [gr-qc]];
A.~De Felice and S.~Tsujikawa,
Phys.\ Rev.\ Lett.\  {\bf 105}, 111301 (2010)
[arXiv:1007.2700 [astro-ph.CO]];
T.~Kobayashi, M.~Yamaguchi and J.~Yokoyama,
Phys.\ Rev.\ Lett.\  {\bf 105}, 231302 (2010)
[arXiv:1008.0603 [hep-th]];
A.~De Felice and S.~Tsujikawa,
Phys.\ Rev.\ D {\bf 84}, 124029 (2011)
[arXiv:1008.4236 [hep-th]];
C.~de Rham and L.~Heisenberg,
Phys.\ Rev.\ D {\bf 84}, 043503 (2011)
[arXiv:1106.3312 [hep-th]];
L.~Heisenberg, R.~Kimura and K.~Yamamoto,
Phys.\ Rev.\ D {\bf 89}, 103008 (2014)
[arXiv:1403.2049 [hep-th]].

\bibitem{Nicolis}
A.~Nicolis, R.~Rattazzi and E.~Trincherini,
Phys.\ Rev.\  D {\bf 79}, 064036 (2009)
[arXiv:0811.2197 [hep-th]].

\bibitem{Ostro}
M.~V.~Ostrogradski, Mem. Acad. St. Petersbourg VI 4,
{\bf 385} (1850);
R.~P.~Woodard,
Lect.\ Notes Phys.\  {\bf 720}, 403 (2007)
[astro-ph/0601672].

\bibitem{Horndeski} 
G.~W.~Horndeski,
Int.\ J.\ Theor.\ Phys.\  {\bf 10}, 363 (1974).

\bibitem{Horn2} 
C.~Deffayet, X.~Gao, D.~A.~Steer and G.~Zahariade,
Phys.\ Rev.\ D {\bf 84}, 064039 (2011)
[arXiv:1103.3260 [hep-th]];
T.~Kobayashi, M.~Yamaguchi and J.~'i.~Yokoyama,
Prog.\ Theor.\ Phys.\  {\bf 126}, 511 (2011)
[arXiv:1105.5723 [hep-th]];
C.~Charmousis, E.~J.~Copeland, A.~Padilla and P.~M.~Saffin,
Phys.\ Rev.\ Lett.\  {\bf 108}, 051101 (2012)
[arXiv:1106.2000 [hep-th]].

\bibitem{GLPV} 
J.~Gleyzes, D.~Langlois, F.~Piazza and F.~Vernizzi,
Phys.\ Rev.\ Lett.\  {\bf 114}, 211101 (2015)
[arXiv:1404.6495 [hep-th]].

\bibitem{building} 
J.~Gleyzes, D.~Langlois, F.~Piazza and F.~Vernizzi,
JCAP {\bf 1308}, 025 (2013)
[arXiv:1304.4840 [hep-th]].

\bibitem{Hami} 
C.~Lin, S.~Mukohyama, R.~Namba and R.~Saitou,
JCAP {\bf 1410}, 071 (2014)
[arXiv:1408.0670 [hep-th]];
X.~Gao,
Phys.\ Rev.\ D {\bf 90}, 104033 (2014)
[arXiv:1409.6708 [gr-qc]];
J.~Gleyzes, D.~Langlois, F.~Piazza and F.~Vernizzi,
JCAP {\bf 1502}, 018 (2015)
[arXiv:1408.1952 [astro-ph.CO]];
C.~Deffayet, G.~Esposito-Farese and D.~A.~Steer,
Phys.\ Rev.\ D {\bf 92}, 084013 (2015)
[arXiv:1506.01974 [gr-qc]].

\bibitem{Gergely} 
L.~A.~Gergely and S.~Tsujikawa,
Phys.\ Rev.\ D {\bf 89}, 064059 (2014)
[arXiv:1402.0553 [hep-th]].

\bibitem{Kase14} 
R.~Kase and S.~Tsujikawa,
Int.\ J.\ Mod.\ Phys.\ D {\bf 23}, no. 13, 1443008 (2015)
[arXiv:1409.1984 [hep-th]].

\bibitem{KT14} 
R.~Kase and S.~Tsujikawa,
Phys.\ Rev.\ D {\bf 90}, 044073 (2014)
[arXiv:1407.0794 [hep-th]].

\bibitem{DKT} 
A.~De Felice, K.~Koyama and S.~Tsujikawa,
JCAP {\bf 1505}, 058 (2015)
[arXiv:1503.06539 [gr-qc]].

\bibitem{conical1} 
A.~De Felice, R.~Kase and S.~Tsujikawa,
Phys.\ Rev.\ D {\bf 92}, 124060 (2015)
[arXiv:1508.06364 [gr-qc]].

\bibitem{conical2} 
R.~Kase, S.~Tsujikawa and A.~De Felice,
JCAP {\bf 1603}, 003 (2016)
[arXiv:1512.06497 [gr-qc]].

\bibitem{Vainshtein} 
A.~I.~Vainshtein,
Phys.\ Lett.\ B {\bf 39}, 393 (1972).

\bibitem{Koba}
T.~Kobayashi, Y.~Watanabe and D.~Yamauchi,
Phys.\ Rev.\ D {\bf 91}, 064013 (2015)
[arXiv:1411.4130 [gr-qc]].

\bibitem{sphe}
R.~Kase, L.~A.~Gergely and S.~Tsujikawa,
Phys.\ Rev.\ D {\bf 90}, 124019 (2014)
[arXiv:1406.2402 [hep-th]];
K.~Koyama and J.~Sakstein,
Phys.\ Rev.\ D {\bf 91}, 124066 (2015)
[arXiv:1502.06872 [astro-ph.CO]];
R.~Saito, D.~Yamauchi, S.~Mizuno, J.~Gleyzes and D.~Langlois,
JCAP {\bf 1506}, 008 (2015)
[arXiv:1503.01448 [gr-qc]].

\bibitem{Babichev} 
E.~Babichev, K.~Koyama, D.~Langlois, R.~Saito and J.~Sakstein,
arXiv:1606.06627 [gr-qc].

\bibitem{KTD}
R.~Kase, S.~Tsujikawa and A.~De Felice,
Phys.\ Rev.\ D {\bf 93}, 024007 (2016)
[arXiv:1510.06853 [gr-qc]].

\bibitem{Barrow}
J.~D.~Barrow, M.~Thorsrud and K.~Yamamoto,
JHEP {\bf 1302}, 146 (2013)
[arXiv:1211.5403 [gr-qc]].

\bibitem{Jimenez1}
J.~Beltran Jimenez and A.~L.~Maroto,
Phys.\ Rev.\ D {\bf 78}, 063005 (2008)
[arXiv:0801.1486 [astro-ph]];
J.~Beltran Jimenez and A.~L.~Maroto,
JCAP {\bf 0903}, 016 (2009)
[arXiv:0811.0566 [astro-ph]];
 J.~B.~Jimenez, R.~Durrer, L.~Heisenberg and M.~Thorsrud,
JCAP {\bf 1310}, 064 (2013)
[arXiv:1308.1867 [hep-th]].

\bibitem{Jimenez2}
  J.~Beltran Jimenez and T.~S.~Koivisto,
  Class.\ Quant.\ Grav.\  {\bf 31}, 135002 (2014)
  [arXiv:1402.1846 [gr-qc]];
J.~Beltran Jimenez and T.~S.~Koivisto,
Phys.\ Lett.\ B {\bf 756}, 400 (2016)
[arXiv:1509.02476 [gr-qc]];
J.~Beltran Jimenez, L.~Heisenberg and T.~S.~Koivisto,
JCAP {\bf 1604}, 046 (2016)
[arXiv:1602.07287 [hep-th]].

\bibitem{TKK}
G.~Tasinato, K.~Koyama and N.~Khosravi,
JCAP {\bf 1311}, 037 (2013)
[arXiv:1307.0077 [hep-th]].

\bibitem{Tasinato}
G.~Tasinato,
JHEP {\bf 1404}, 067 (2014)
[arXiv:1402.6450 [hep-th]];
G.~Tasinato,
Class.\ Quant.\ Grav.\  {\bf 31}, 225004 (2014)
[arXiv:1404.4883 [hep-th]]. 

\bibitem{Fleury}
P.~Fleury, J.~P.~B.~Almeida, C.~Pitrou and J.~P.~Uzan,
JCAP {\bf 1411}, 043 (2014).
[arXiv:1406.6254 [hep-th]].

\bibitem{Hull}
M.~Hull, K.~Koyama and G.~Tasinato,
JHEP {\bf 1503}, 154 (2015)
[arXiv:1408.6871 [hep-th]];
M.~Hull, K.~Koyama and G.~Tasinato,
Phys.\ Rev.\ D {\bf 93}, 064012 (2016)
[arXiv:1510.07029 [hep-th]].


\bibitem{Lagos:2016wyv} 
  M.~Lagos, T.~Baker, P.~G.~Ferreira and J.~Noller,
  JCAP {\bf 1608}, no. 08, 007 (2016)
  [arXiv:1604.01396 [gr-qc]].



\bibitem{Heisenberg} 
L.~Heisenberg,
JCAP {\bf 1405}, 015 (2014)
[arXiv:1402.7026 [hep-th]].

 
\bibitem{Allys}
E.~Allys, P.~Peter and Y.~Rodriguez,
JCAP {\bf 1602}, 004 (2016)
[arXiv:1511.03101 [hep-th]];
E.~Allys, J.~P.~B.~Almeida, P.~Peter and Y.~Rodriguez,
arXiv:1605.08355 [hep-th].
 
\bibitem{Jimenez2016} 
J.~B.~Jimenez and L.~Heisenberg,
Phys.\ Lett.\ B {\bf 757}, 405 (2016)
[arXiv:1602.03410 [hep-th]].

\bibitem{Cosmo} 
A.~De Felice, L.~Heisenberg, R.~Kase, S.~Mukohyama, S.~Tsujikawa and Y.~l.~Zhang,
JCAP {\bf 1606}, 048 (2016)
[arXiv:1603.05806 [gr-qc]].

\bibitem{Geff} 
A.~De Felice, L.~Heisenberg, R.~Kase, S.~Mukohyama, S.~Tsujikawa and Y.~l.~Zhang,
Phys.\ Rev.\ D {\bf 94}, 044024 (2016)
[arXiv:1605.05066 [gr-qc]].

\bibitem{screening} 
A.~De Felice, L.~Heisenberg, R.~Kase, S.~Tsujikawa, Y.~l.~Zhang and G.~B.~Zhao,
Phys.\ Rev.\ D {\bf 93}, 104016 (2016)
[arXiv:1602.00371 [gr-qc]].

\bibitem{Minami} 
M.~Minamitsuji,
arXiv:1607.06278 [gr-qc].

\bibitem{Niz} 
J.~Chagoya, G.~Niz and G.~Tasinato,
Class.\ Quant.\ Grav.\  {\bf 33}, no. 17, 175007 (2016)
[arXiv:1602.08697 [hep-th]].

\bibitem{HKT} 
L.~Heisenberg, R.~Kase and S.~Tsujikawa,
Phys.\ Lett.\ B {\bf 760}, 617 (2016)
[arXiv:1605.05565 [hep-th]].

\bibitem{anisotropy} 
L.~Heisenberg, R.~Kase and S.~Tsujikawa,
arXiv:1607.03175 [gr-qc].

\bibitem{KNY} 
R.~Kimura, A.~Naruko and D.~Yoshida,
arXiv:1608.07066 [gr-qc].

\bibitem{scava}
C.~Burrage and D.~Seery,
JCAP {\bf 1008}, 011 (2010).
[arXiv:1005.1927 [astro-ph.CO]];
P.~Brax, C.~Burrage and A.~C.~Davis,
JCAP {\bf 1109}, 020 (2011)
[arXiv:1106.1573 [hep-ph]];
A.~De Felice, R.~Kase and S.~Tsujikawa,
Phys.\ Rev.\ D {\bf 85}, 044059 (2012).
[arXiv:1111.5090 [gr-qc]];
R.~Kimura, T.~Kobayashi and K.~Yamamoto,
Phys.\ Rev.\ D {\bf 85}, 024023 (2012).
[arXiv:1111.6749 [astro-ph.CO]];
K.~Koyama, G.~Niz and G.~Tasinato,
Phys.\ Rev.\ D {\bf 88}, 021502 (2013)
[arXiv:1305.0279 [hep-th]];
R.~Kase and S.~Tsujikawa,
JCAP {\bf 1308}, 054 (2013)
[arXiv:1306.6401 [gr-qc]].

\bibitem{Will} 
C.~M.~Will,
Living Rev.\ Rel.\  {\bf 9}, 3 (2006)
[gr-qc/0510072].

\end{thebibliography}
\end{document}